\documentclass{iopart}
\usepackage{graphicx}
\usepackage{iopams}
\usepackage{hyperref}
\usepackage{units}
\usepackage{multirow}

\newcommand{\vG}{\ve{G}}

\newcommand{\abso}[1]{\vert #1 \vert}
\newcommand{\ve}[1]{\ensuremath{\boldsymbol{#1}}}
\newcommand{\aver}[1]{\langle #1 \rangle}
\newcommand{\cre}[2]{\ensuremath{#1_{#2}^\da}}
\newcommand{\ann}[2]{\ensuremath{#1_{#2}^{ }}}
\newcommand{\da}{\dagger}

\begin{document}

\title[Electronic spectra of DWNTS in parallel magnetic field]{Electronic spectra of commensurate and incommensurate DWNTs in parallel magnetic field}
\author{Magdalena Marga\'{n}ska, Shidong Wang and Milena Grifoni}
\address{Institut f\"{u}r Theoretische Physik, Universit\"at Regensburg, 93040 Germany}

\date{\today}

\begin{abstract}
We study the electronic spectra of commensurate and incommensurate double-wall carbon nanotubes (DWNTs) of finite length. 
The coupling between nanotube shells is taken into account as an intershell electron tunneling.
Selection rules for the intershell coupling are derived.
Due to the finite size of the system, these rules do not represent exact conservation of the crystal momentum, but only an approximate one; therefore the coupling between longitudinal momentum states in incommensurate DWNTs becomes possible. 
The use of the selection rules allows a fast and efficient calculation of the electronic spectrum. In the presence of a magnetic field parallel to the DWNT axis we find spectrum modulations which depend on the chiralities of the shells.
\end{abstract}

\maketitle

\section{Introduction}
\label{sec:introduction}

Due to their unusual physical properties, cf. e.g.~\cite{saito:1998,book:2006}, carbon nanotubes have become promising building blocks for nanotechnology applications and have attracted a lot of attention since their discovery. 
Carbon nanotubes can be single-walled (SWNT) or multi-walled (MWNT), depending on whether they consist of a single or of several graphene sheets wrapped onto coaxial cylinders (so called ``shells''), respectively. Electronic properties of SWNTs have been mostly understood~\cite{saito:1998}. For example, SWNTs are usually ballistic conductors~\cite{white:nature1998}, and whether a SWNT is metallic or semiconducting is solely determined by its geometry.  
However, the situation is much less clear for MWNTs. In fact, due to the additional shells, MWNTs exhibit qualitatively different properties than SWNTs. Except for few experiments, see e.g.~\cite{frank:science1998, urbina:prl2003}, MWNTs are typically diffusive conductors~\cite{langer:prl1996, bachtold:nature1999}, with current being carried by the outermost shell at low bias~\cite{bachtold:nature1999, fujiwara:prb1999} and
also by inner shells at high bias~\cite{collins:prl2001}. A recent experiment reported that the intershell conductance is quite weak and consistent with the tunneling through the orbitals of nearby
shells~\cite{bourlon:prl2004}. The difficulty in a theoretical description of MWNTs lies in the fact
that the coaxial shells have usually different chiralities. In such case MWNTs are intrinsically aperiodic, since a common unit cell for the whole object cannot be defined due to the respective symmetries of individual shells.\\
The simplest system in which the inter-shell effects can be studied is a double-walled nanotube (DWNT). 
The two shells are coupled by weak van der Waals interactions, which give rise to an inter-shell electron tunneling. DWNTs have been studied in various approaches. 
By using ab-initio methods on graphite, effective inter-layer hopping integrals have been found ~\cite{charlier:prb1991}, closely matching experimental results \cite{misu:jpsj1979,ohta:prl2007}. In the calculations involving nanotubes the hopping parameters are usually considered to be similar to those in graphite.
An ab-initio study of multiwall nanotubes \cite{charlier:prl1993} correctly predicted the intershell distance and freedom of telescopic and rotational motion of the shells, later confirmed experimentally \cite{zettl:science2000}. Commensurate DWNTs have been thoroughly analyzed, and their electronic spectra \cite{saito:jap1993,kwon:prb1998,lin:physicae2006,pudlak:condmat2007} and transport properties \cite{triozon:prb2004} have been discussed. Some authors investigated also the properties of incommensurate DWNTs, looking at spectral correlations \cite{ahn:prl2003,uryu:prb2004} and transport properties \cite{uryu:prb2004,wang:prl2005,yoon:prb2002,lunde:prb2005,wang:2006} or simulating their STM images \cite{lambin:prb2000}. 
Each of the transport studies refers in some way to the selection rules for the inter-shell coupling. They are mentioned in passing in ~\cite{uryu:prb2004} when discussing the tunneling between states in the inner and outer shell at Fermi points $K$ and $K'$. The general analysis of MWNT conduction presented in ~\cite{yoon:prb2002} relies on the conservation of quasi-crystal momentum to prove that the conductance of a long MWNT is dominated by the outermost shell. In \cite{lunde:prb2005} the authors consider a long DWNT and calculate the intershell resistance, as coming only from the Coulomb drag, i.e. neglecting the inter-shell tunneling. They find selection rules for the coupling between momentum states in different shells. Since the interest in the above works is focused on the conduction, they explore the consequences of those rules mainly close to the Fermi level.\\
When a uniform magnetic field is applied to a system, interesting and subtle effects occur, depending on the geometry and topology of the system, due to the {\em Peierls phase}~\cite{peierls:zphys1933} acquired by the electronic wavefunction. For electrons moving in spatially periodic potentials, if the flux through the elementary cell contains an irrational number of flux quanta, the periodicity is destroyed and the spectrum becomes fractal ~\cite{hofstadter:prb1976}. When the field is applied parallel to the axis of symmetry of a ring or cylinder, it causes the Aharonov-Bohm effect or persistent currents ~\cite{buttiker:prb1985,gefen:prb1988}. In nanotubes, because of their unique dispersion relation, the field can induce e.g. a periodic metal-semiconductor transition, predicted in \cite{ajiki:jpsj1993a} and observed in many experiments \cite{mceuen:nature2004,coskun:science2004,strunk:sst2006,lassagne:prl2007}.
The effects of a uniform magnetic field on the spectrum of a commensurate DWNT in the vicinity of the Fermi level have also been studied, taking into account several rotational configurations of the two shells \cite{latge:carbon2007,lin:jpcm2008}. The tunneling coupling between shells of a DWNT modifies the spectra of the individual shells, introducing numerous avoided crossings, which in turn result in the depletion of the density of states (DOS) in one or more regions of the spectrum~\cite{nemec:prb2006}. In small fields this region lies close to the bottom of the valence band, but when the magnetic field increases, the influence of the intershell coupling is visible in higher energy ranges. \\
In this work we extend to finite size DWNTs an approach presented in ~\cite{wang:prl2005} in which the Hamiltonian of DWNTs is analyzed in the reciprocal space.  We find the selection rules for the coupling between momentum states and estimate the amplitude of the coupling.
This method has the advantage of being computationally fast, due to the action of the selection rules and can be applied to {\em commensurate} as well as {\em incommensurate} DWNTs. For short DWNTs both our method and the direct diagonalization of the tight-binding Hamiltonian in the real space yield spectra with the same positions of the van Hove peaks with some mismatches in their heights.
For commensurate DWNTs in parallel magnetic field our result matches the results of ~\cite{nemec:prb2006}, where a similar system has been studied in the real space.
We also calculate the electronic spectra in changing magnetic field for incommensurate DWNTs. We find a periodic closing and opening of the gap at the Fermi level, as well as a region with depleted density of states (DOS). This region evolves with the magnetic field in a complex way, determined by the geometry of the two shells.\\
This paper is organized as follows. In \sref{sec:lattice} we introduce various quantities needed for the characterization of the real and reciprocal space of graphene and of nanotubes. The intershell tunneling in DWNTs is studied in \sref{sec:coupling}, where the reciprocal space formula for the tunneling coupling is derived and analyzed. The coupling changes when a uniform magnetic field is applied and its influence on the energy spectrum is studied in \sref{sec:magnetic}. \Sref{sec:results} concludes this work.

\section{Direct and reciprocal lattice structure of DWNTs}
\label{sec:lattice}

\subsection{Graphene}
\begin{figure}[htpb]
\begin{center}
\includegraphics[width=3.5cm,angle=-90]{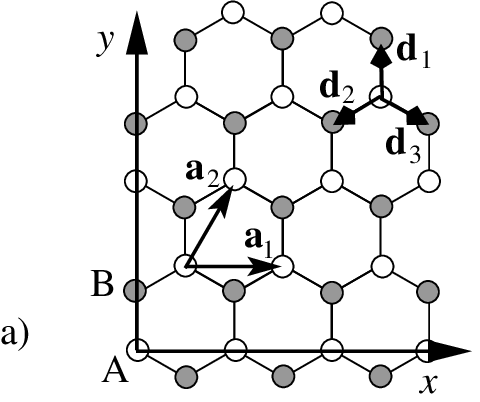}\hspace{.5cm}
\includegraphics[width=4cm,angle=-90]{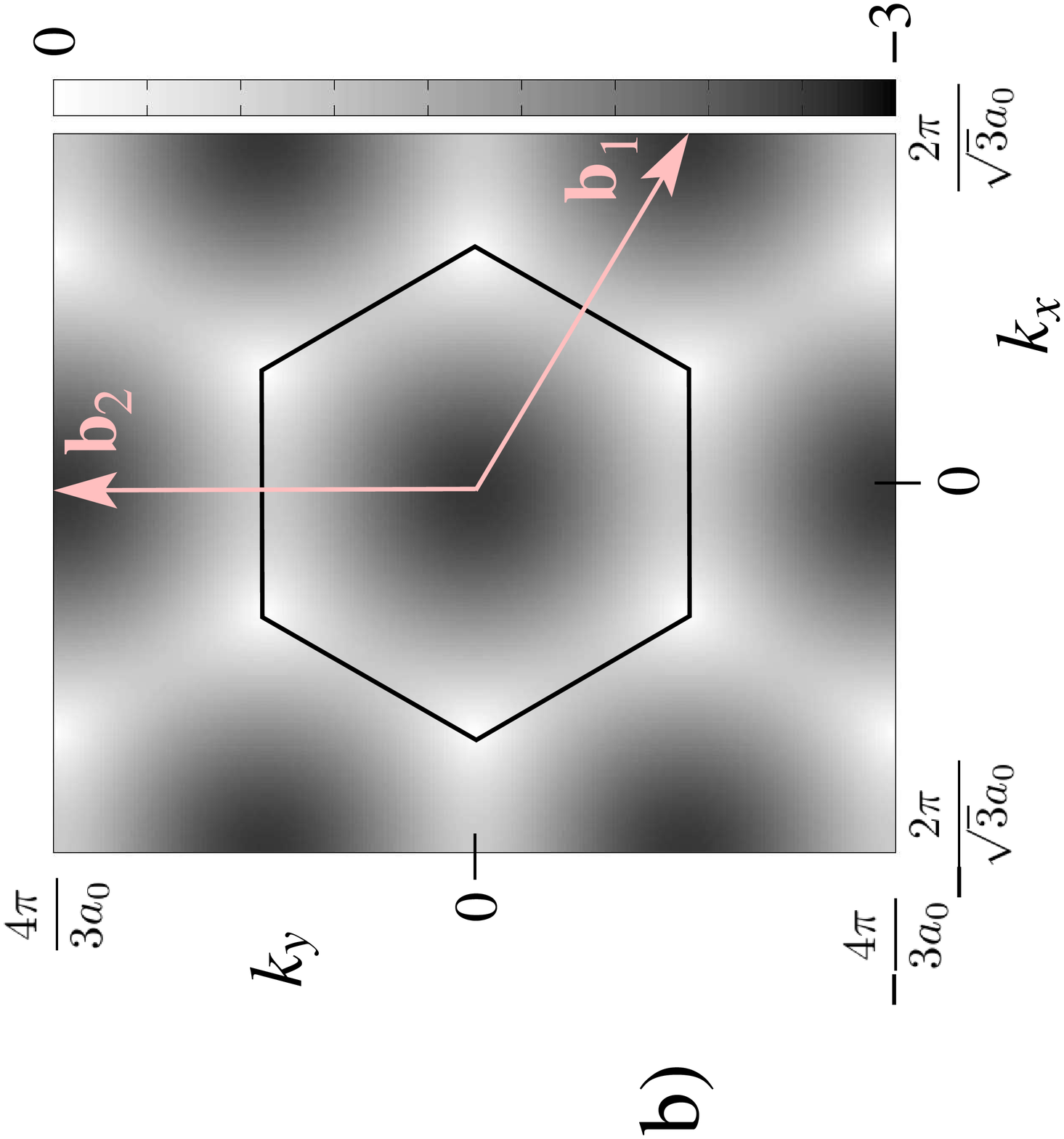}
\end{center}
\caption{Direct and reciprocal lattice of graphene. a) Atomic structure of the honeycomb lattice with two sublattices $A$ and $B$ and the lattice generators $\ve{a}_1,\ve{a}_2$. Vectors $\ve{d}_i$ connect the atoms from sublattice $A$ with their nearest neighbours. b) The first Brillouin zone of graphene and the reciprocal lattice generators $\ve{b}_1,\ve{b}_2$. The background is a greyscale map of the negative part of the dispersion relation in $\gamma_0$ units.}
\label{fig:graphene}
\end{figure}

Various nanostructures, such as nanotubes, graphene ribbons or nanocones \cite{charlier:prl2001,chen:apl2006} can be treated as fragments of a graphene sheet (\fref{fig:graphene}) with appropriate boundary conditions. For later purposes we briefly recapitulate how to characterize the graphene lattice and its electronic spectrum and how to adapt this description to the case of carbon nanotubes. The honeycomb lattice of graphene is generated by two basis vectors of equal length, and the angle between them is 60$^\circ$. We choose their Cartesian coordinates as 
\begin{equation}
\ve{a}_1=(\sqrt{3}a_0,0), \hspace{.5cm} \ve{a}_2=\left(\frac{\sqrt{3}}{2}a_0,\frac{3}{2}a_0\right),
\end{equation}
where $a_0=1.42${\AA} is the length of a $C-C$ bond. The elementary cell contains two atoms which generate the two sublattices of graphene through the translations by multiples of $\ve{a}_1,\ve{a}_2$. The atoms $A$ and $B$ in the elementary cell are shifted with respect to the origin of coordinates by vectors $\ve{\tau}_A, \ve{\tau}_B$, respectively. We will refer to these vectors as the {\em sublattice shifts}. In the Cartesian coordinates which we have chosen, they are given by 
\begin{equation}
 \ve{\tau}_A = (0,0), \hspace{.5cm} \ve{\tau}_B = (0,a_0).
\end{equation}
The generators of the reciprocal lattice are 
\begin{equation}
\label{eq:reciprocal-generators}
\ve{b}_1=\left( \frac{2\pi}{\sqrt{3}a_0}, -\frac{2\pi}{3a_0}\right),\hspace{.5cm} 
\ve{b}_2=\left(0,\frac{4\pi}{3a_0}\right). 
\end{equation}
A common starting point for the calculation of the band structure of graphene is the tight-binding model for noninteracting $p_z$ electrons~\cite{saito:1998}, described by the Hamiltonian
\begin{equation}
\label{eq:tb}
H = \sum_{\aver{ij}} \gamma_0 \cre{c}{i\sigma} \ann{c}{j\sigma} \;, 
\end{equation}
where $\gamma_0 \sim -2.7$eV is the hopping integral in graphene; $i,j$ are the $p_z$ orbitals of carbon atoms at positions $i$ and $j$, respectively; $\sigma$ denotes the electron spin and the sum runs over nearest neighbours in the real space. 
The dispersion relation can be derived (see  \sref{sec:coupling} for details). It reads
\begin{equation}
\label{eq:dispersion}
\fl \varepsilon_{\nu}(\ve{k}) = \nu \gamma_0 \sqrt{3 + 2\cos(\ve{k}\cdot\ve{a}_1) +
   2\cos(\ve{k}\cdot\ve{a}_2) + 2\cos(\ve{k}\cdot(\ve{a}_2-\ve{a}_1))},
\end{equation}
where $\nu = +1$ in the conduction band and $\nu=-1$ in the valence band.
This dispersion relation has the characteristic shape of a double crown, with six Fermi points -- only two of them being geometrically inequivalent. A fragment of the atomic lattice of graphene and its reciprocal lattice are shown in \fref{fig:graphene}.\\
\subsection{Single-wall nanotube (SWNT)}
\label{sec:lattice-swnt}
\begin{figure}[htbp]
\begin{center}
\includegraphics[width=3.8cm,angle=-90]{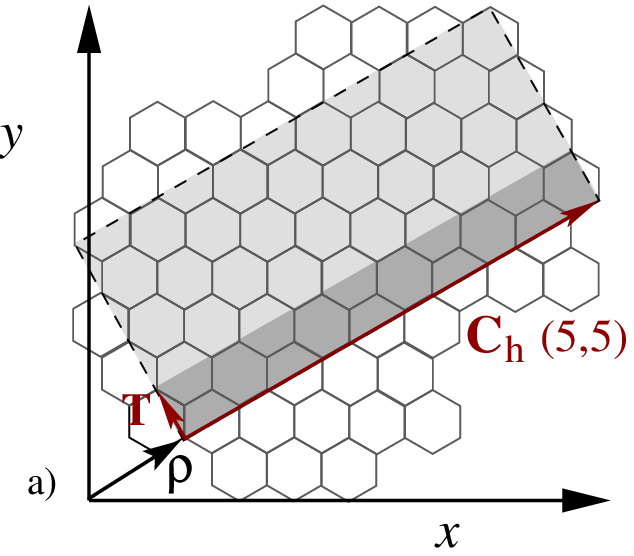}
\includegraphics[width=3.8cm,angle=-90]{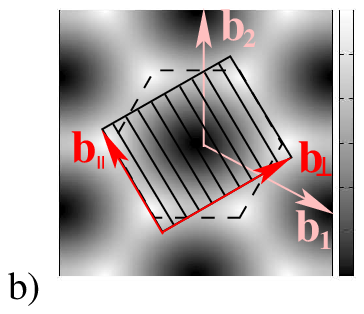}
\includegraphics[width=3.8cm,angle=-90]{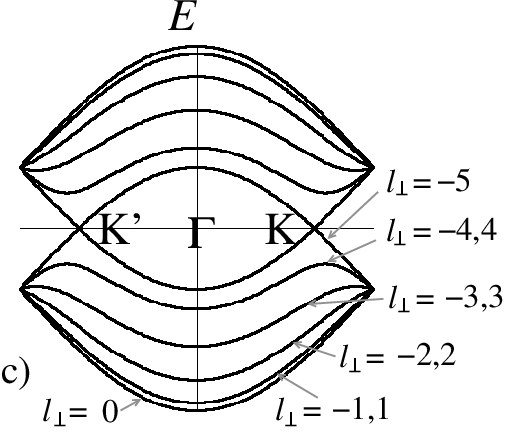}
\end{center}
\caption{ Characterization of an armchair nanotube. a) Unrolled nanotube patch (light grey) on a graphene lattice -- the chiral vector is (5,5) and the nanotube has only four unit cells. The area of the unit cell, spanned by vectors $\ve{C}_h$  and $\ve{T}$ is marked in dark grey. b) The Brillouin zone of graphene (dashed lines) and the {\em reciprocal cell} (solid lines) of an infinite (5,5) tube with the allowed momentum states. c) The electronic subbands of an infinite (5,5) nanotube. Quantum numbers of the subbands in the conduction band $E>0$ are the same as their equivalents in the valence band.}
\label{fig:swnt}
\end{figure}

A single-wall nanotube can be described as a rectangular patch of graphene with two opposite sides joined together by periodic boundary conditions (\fref{fig:swnt}a). The vector defining the circumference of the SWNT is called {\em chiral vector} and is uniquely defined by two coordinates in the basis of lattice generators
\begin{equation}
\label{eq:Ch}
 \ve{C}_h = m_1 \ve{a}_1 + m_2 \ve{a}_2,\hspace{.5cm} m_1,m_2 \in\mathbb{Z}.
\end{equation}
Because of the hexagonal symmetry of graphene this notation is redundant. In particular, the nanotube with $(-m_1,-m_2)$ is identical to the one with $(m_1,m_2)$, and $(m_2,m_1)$ is its mirror image. The convention is to keep $m_1\geq m_2$ and $m_2\geq 0$. In most nanotubes a chiral arrangement of atoms can be observed along the nanotube. There are only two combinations of parameters which describe achiral nanotubes: $(m,0)$ corresponding to so-called zigzag tubes and $(m,m)$ corresponding to armchair tubes. The nanotubes can also be viewed as objects created by a repeated translation of a unit cell, defined by the vectors $\ve{C}_h$ and
$\ve{T}$ (see \fref{fig:swnt}a):
\begin{equation}
\ve{T} = -\frac{m_1+2m_2}{d_R} \ve{a}_1 + \frac{2m_1+m_2}{d_R} \ve{a}_2. 
\end{equation}
Here $d_R$ is the greatest common divisor of $(m_1+2m_2)$ and $(2m_1+m_2)$. \\
The boundary conditions around the circumference of the nanotube (in transverse direction) are always periodic ({\em PBC}). There are two ways of dealing with the 
boundary conditions along the nanotube axis (in the longitudinal direction), resulting in the same spectrum. One way is to consider {\em open} boundary conditions ({\em OBC}) with the wavefunctions defined on the length of the nanotube. The other way is to consider periodic boundary conditions on a nanotube twice that length and to choose only the energy eigenfunctions which are antisymmetric with respect to the center of the extended tube. Physically it means that we choose only those wavefunctions which are reflected from the end of the original tube (or the center of the extended tube) with opposite phase. This restriction removes both the level degeneracy caused by {\em PBC} and the $k_\parallel=0$ eigenstate, which is symmetric with respect to the center of the extended tube. \\
The boundary conditions cause the quantization of momentum
\begin{equation}
\label{eq:quantization}
 \ve{k} = (k_\perp, k_\parallel) = \left( \frac{2\pi}{C_h} l_\perp, \frac{\pi}{L} l_\parallel\right),
\hspace{.5cm} l_\perp,l_\parallel\in\mathbb{Z},
\end{equation}
where $L = M \vert \ve{T} \vert$ is the length of the nanotube, equal to $M$ unit cells. Note that in \eref{eq:quantization} open boundary conditions along the nanotube axis have been assumed.
In infinite nanotubes $k_\parallel$ is continuous and the allowed momentum states are a set of lines of constant $k_\perp$. Instead of working in the quantized hexagonal Brillouin zone of graphene, it is more comfortable to define a rectangular unit cell of the reciprocal space, with the area equal to that of the Brillouin zone and yielding the same energy spectrum (\fref{fig:swnt}b). We shall refer to it as the {\em reciprocal cell}. It is spanned by vectors $\ve{b}_\perp$ and $\ve{b}_\parallel$ given in the basis of graphene reciprocal lattice generators by
\numparts
\begin{eqnarray}
\label{eq:bperp}
 & \ve{b}_\perp & =  \frac{2m_1+m_2}{d_R} \ve{b}_1 + \frac{m_1+2m_2}{d_R} \ve{b}_2 , \\
 & \ve{b}_\parallel & =  -\frac{m_2}{S}\ve{b}_1 +  \frac{m_1}{S} \ve{b}_2.
\label{eq:bparallel}
\end{eqnarray}
\endnumparts
Notice that the coordinates of $\ve{b}_\perp$ are integer, therefore $\ve{b}_\perp$ is always a reciprocal lattice vector. Note also that, since $l_\perp = k_\perp R$, the angular momentum is $\hbar\, l_\perp$. \\
The projection of the lines of constant $k_\perp$ on the dispersion relation reduces the full 2D spectrum to a set of 1D subbands, numbered by their value of angular momentum quantum number $l_\perp$ (\fref{fig:swnt}c). The number $S$ of subbands in one band, equal to the number of allowed values of $l_\perp$, is the number of graphene unit cells in the unit cell of the nanotube
\begin{equation}
 S(m_1,m_2) = \frac{2(m_1^2 + m_1 m_2 + m_2^2)}{d_R}.
\end{equation}
Each subband has a positive and a negative energy branch, accounting for the presence of two atoms in the graphene unit cell. In the reciprocal cell all subbands contain equal number of $k_\parallel$ states. In finite nanotubes the 1D subbands are further discretized, and a nanotube containing $M$ unit cells has $M$ longitudinal momentum values in each of the $S$ subbands. Therefore $0 < l_\parallel \leq M$ and the allowed range of $l_\perp$ is $\left[ -S/2, S/2 \right)$.
\subsection{Double-wall nanotube (DWNT)}
\begin{figure}[htpb]
\begin{center}
\includegraphics[width=4cm,angle=-90]{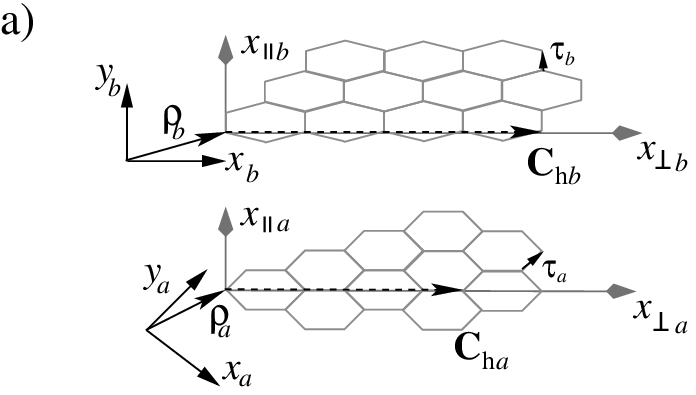}
\includegraphics[width=4cm,angle=-90]{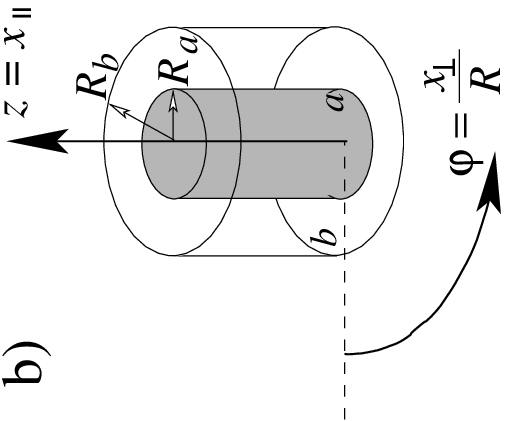}
\end{center}
\caption{Systems of coordinates used to describe a location on a double-walled nanotube. a) Two graphene layers. The vectors $\brho_a, \brho_b$ describing the relative position of the graphene patches of the two shells (armchair $a$ and zigzag $b$) and shifts $\btau_a, \btau_b$ between $A$ and $B$ sublattices in both shells are indicated. b) Schematic view of a DWNT and its system of coordinates $(x_\perp, x_\parallel)$. }
\label{fig:dwnt}
\end{figure}

A double-wall nanotube consists of two coaxial single-wall nanotubes, called also {\em shells}. The inter-shell distance $\Delta$ is typically of the order of 3.4\AA ~\cite{ijima:nature1991}. The coupling between two shells can be taken into account as an inter-shell tunneling of electrons. The implications of this tunneling will be explored in \sref{sec:coupling}. A schematic picture of a DWNT and its system of coordinates is shown in \fref{fig:dwnt}.\\
In the present paper we will be using several systems of coordinates, each of them suitable for a particular purpose. For the first three systems of coordinates we start from two graphene layers separated by a distance $\vert R_b - R_a \vert$.
Each point on one of the constituent 2D graphene layers of a DWNT can be described either by the Cartesian coordinates $(x,y)$ or by the nanotube patch coordinates $(x_\perp, x_\parallel)$. The third possibility is the system defined by $(\ve{a}_1,\ve{a}_2)$, but this one is used only in the definition of the nanotube chirality $\ve{C}_h$. When the nanotube is rolled, it becomes a 3D object and the most natural coordinate system is the cylindrical one. The cylindrical coordinates $(r,\varphi,z)$ of a point on the shell $\beta$ are related to the 2D nanotube coordinates $(x_\perp,x_\parallel)$ by 
\begin{equation}
(r,\varphi,z)_\beta = (R_\beta, \frac{x_{\perp \beta}}{R_\beta},x_{\parallel \beta}).
\end{equation}
In the reciprocal space we use only the 2D coordinates of the graphene layers. 
The vectors $\ve{G}$ in the reciprocal space can be expressed as $(n_1,n_2)$ in the graphene basis of $(\ve{b}_1,\ve{b}_2)$, i.e. $\ve{G} = n_1 \ve{b}_1 + n_2 \ve{b}_2$, or as $(G_\perp, G_\parallel)$ in the basis of $(\ve{b}_\perp, \ve{b}_\parallel)$ spanning the reciprocal cell of a nanotube (see \eref{eq:bperp} and \eref{eq:bparallel}). \\

\section{Effective intershell coupling in DWNTs}
\label{sec:coupling}
The starting point for our investigation of the consequences of the inter-shell electron tunneling is a tight-binding model for noninteracting $p_z$ electrons on each shell of the carbon nanotube~\cite{saito:1998}. The tight-binding Hamiltonian of a DWNT is obtained from that of two graphene sheets $a$ and $b$ placed on top of each other at a distance $\abso{R_a - R_b}$ by imposing periodic boundary conditions along the directions determined by the chiral vectors $\ve{C}_{ha} =(m_{1a}, m_{2a})$ and $\ve{C}_{hb}=(m_{1b}, m_{2b})$ (see \eref{eq:Ch}). The DWNT Hamiltonian  is
\begin{equation}
\label{eq:H-total}
  H = H_0 + H_t =  \sum_{\beta\sigma} \sum_{\aver{ij}} \gamma_0 \cre{c}{\beta i\sigma}
  \ann{c}{\beta j\sigma}  + 
  \sum_{ij\sigma} t_{\ve{r}_{ai}, \ve{r}_{bj}} \cre{c}{ai\sigma} \ann{c}{bj\sigma}
+ \mathrm{H.c.}, 
\end{equation}
where the operators $\cre{c}{\beta j\sigma}$ and $\ann{c}{\beta j\sigma}$ are creation and annihilation operators of an electron with spin $\sigma$ on shell $\beta$ at site $j$, respectively. Here $\beta=a,b$ is the shell index and, as in \eref{eq:tb}, $\aver{ij}$ is a sum over nearest neighbors and $\gamma_0 \sim \unit[-2.7]{eV}$ is the intrashell nearest neighbor coupling. The spin-independent intershell coupling $t_{ai, bj}$ is assumed to
depend exponentially on the distance between two atoms, $d(\ve{r}_{ai}, \ve{r}_{bj})$, as
\begin{equation}
  \label{eq:intershell-coupling}
   t_{\ve{r}_{ai}, \ve{r}_{bj}} = t_0\cos\theta_{ij} e^{-(d(\ve{r}_{ai},
    \ve{r}_{bj}) - \Delta)/a_t},
\end{equation}
where $t_0 \sim \unit[-0.34]{eV}$, $\Delta \sim \unit[0.34]{nm}$, $\theta_{ij}$ is
the angle between the $p_z$-orbitals of the two atoms, and $a_t \sim 0.45${\AA}~\cite{saito:1998}
is a parameter controlling the range of the tunneling. We adopt here the second approach to the boundary conditions along the nanotube axis, described in \sref{sec:lattice-swnt}. We extend our DWNT to twice its original length, assume periodic boundary conditions, and reject all solutions which are symmetric with respect to the center of the extended nanotube. The sum over $i,j$ runs therefore over the extended nanotube.\\
 It is convenient to express the Hamiltonian in the basis of plane waves in each individual shell~\cite{saito:1998, maarouf:prb2000}. We introduce the electron operators
\begin{equation*}
\fl  \ann{c}{\beta j \sigma} = \frac{1}{\sqrt{2N_\beta}}\sum_{\ve{k}} e^{i \ve{k} \cdot \ve{r}_j}
  \ann{c}{\beta p(j) \ve{k} \sigma}, \qquad
  \cre{c}{\beta j \sigma} = \frac{1}{\sqrt{2N_\beta}}\sum_{\ve{k}} e^{-i \ve{k} \cdot \ve{r}_j}
  \cre{c}{\beta p(j) \ve{k} \sigma},
\end{equation*}
where $p=A,B$ is the index for the two interpenetrating sublattices in a graphene sheet, and $N_\beta$ is the number of graphene unit cells on shell $\beta$. The extended tube has twice as many atoms as the original one, hence the $\sqrt{2}$ in the normalization factor. The Hamiltonian takes the form
\begin{equation}
 \fl H =  
\sum_{\beta p \ve{k}\sigma} \gamma_{\ve{k}}
 \cre{c}{\beta p \ve{k}\sigma} \ann{c}{\beta p' \ve{k}\sigma}
   +
  \sum_{\ve{k}_a\ve{k}_{b}} \sum_{p_ap_{b}\sigma}
 \mathcal{T}_{p_ap_{b}}(\ve{k}_a,\ve{k}_{b})\cre{c}{ap_a\ve{k}_a\sigma}
  \ann{c}{bp_{b}\ve{k}_{b}\sigma} + \mathrm{H.c.}  \;,
\label{eq:H-plane}
\end{equation}
where the {\em intra}shell coupling is $\gamma_{\ve{k}} = \sum_{j=1}^3 \gamma_0 e^{i\ve{k}\cdot\ve{d}_j}$, with $\ve{d}_j$ the vectors connecting an $A$ sublattice atom to its three nearest neighbours in sublattice $B$ (\fref{fig:graphene}a). The position of each atom in the graphene patch can be expressed as
$\ve{r}_{\beta}=\ve{R}+\ve{X}_{\beta}$, with $\ve{R}$ a graphene lattice vector, $\ve{X}_{\beta}=\brho_\beta+\btau_{\beta p}$, where $\brho_a - \brho_b$ describes the relative position of
the two shells and $\btau_{\beta p}$ is the appropriate sublattice shift, cf. \fref{fig:dwnt}. The elements of the {\em  inter}shell $2\times 2$ coupling matrix can be expressed as~\cite{maarouf:prb2000}
\begin{equation}
  \label{eq:effective-coupling}
\mathcal{T}_{p_ap_b}(\ve{k}_a, \ve{k}_b) =
  \sum_{\ve{G}_a\ve{G}_b}
    e^{i\ve{G}_a\cdot \ve{X}_a  -
    i\ve{G}_b\cdot \ve{X}_b} t_{\ve{k}_a+\ve{G}_a,
    \ve{k}_b+\ve{G}_b}\;.
\end{equation}
Here $\ve{G}$ is the graphene reciprocal lattice vector $\vG = n_1\ve{b}_1 + n_2\ve{b}_2\equiv (n_1,n_2)$. The intershell coupling has the form 
\begin{equation}
\label{eq:t-k}
 t_{\ve{q}_a, \ve{q}_b} =
    \frac{1}{A_{\mathrm{cell}}^2\sqrt{4N_aN_b}} \int d\ve{r}_a d\ve{r}_b
    e^{i (\ve{q}_b\cdot\ve{r}_b - \ve{q}_a\cdot\ve{r}_a)}
    t_{\ve{r}_a, \ve{r}_b},
\end{equation}
with $A_{\mathrm{cell}}$ the area of a graphene unit cell and the integral taken over the area of the system, in our case over the extended nanotube.
For the purpose of calculating the energy spectrum, it is better to use the basis of the eigenstates (Bloch states) of the Hamiltonian \eref{eq:H-plane} in the absence of intershell coupling. This can be achieved by the unitary transformation
\begin{equation}
\label{eq:basis-change}
U=\frac{1}{\sqrt 2}\left(
\begin{array}{cc} 
   \frac{\gamma_{\ve{k}}}{\abso{\gamma_{\ve{k}}}} &  
   -\frac{\gamma_{\ve{k}}}{\abso{\gamma_{\ve{k}}}} \\
  1 & 1
\end{array}
\right). 
\end{equation}
The tunneling matrix elements between two Bloch states in different shells can be obtained as
\begin{equation}
  \label{eq:t-bloch}
\tilde{\mathcal{T}}_{\nu_a\nu_b} = (U^\dagger {\cal T}U)_{\nu_a\nu_b}.
\end{equation}
Here $\nu=\mp$ is the index for two graphene bands corresponding to negative/positive energies $\varepsilon_{\beta,\nu}(\ve{k})$ with $\beta=a,b$, where the dispersion relation of these bands is,
cf. \eref{eq:dispersion},
\begin{equation*}
\varepsilon_{\beta,\nu}(\ve{k}) = \nu \gamma_0 \sqrt{3 + 2\cos(\ve{k}\cdot\ve{a}_1) +
   2\cos(\ve{k}\cdot\ve{a}_2) + 2\cos(\ve{k}\cdot(\ve{a}_2-\ve{a}_1))}.
\end{equation*}
The electronic momenta are quantized according to the boundary conditions 
\begin{equation}
\label{eq:bc}
\ve{k}_\beta\cdot\ve{C}_{h\beta} = \frac{2\pi}{C_{h\beta}} l_{\perp \beta},\hspace{.5cm} 
\ve{k}_\beta\cdot 2\ve{L}_\beta = \frac{2\pi}{2L_\beta} l_{\parallel \beta}, \hspace{.75cm} l_{\perp \beta}, l_{\parallel \beta} \in \mathbb{Z}.
\end{equation}
In order to calculate the inter-shell coupling \eref{eq:t-k} we shall use nanotube coordinates,
$(R, x_\perp/R, x_\parallel)$.  The distance between two atoms $a$ and $b$ with cylindrical coordinates $(R_a,\varphi_a,z_a)$ and $(R_b,\varphi_b, z_b)$ is thus
\begin{eqnarray*}
\fl d(\ve{r}_a , \ve{r}_b)
& \equiv D\left( \frac{x_{\perp b}}{R_b} - \frac{x_{\perp a}}{R_a}, x_{\parallel b} - x_{\parallel a}\right) \\
\fl & =  \sqrt{ \abso{R_a - R_b}^2 + 4R_aR_b
   \sin^2\left[\frac{1}{2} \left( \frac{x_{\perp b}}{R_b} - \frac{x_{\perp a}}{R_a}\right) \right] +
 ( x_{\parallel b} - x_{\parallel a} )^2}.
\end{eqnarray*}
For our value of the parameter $a_t$, $\cos\theta_{ij}\approx 1$. Given the form \eref{eq:intershell-coupling}, the
intershell coupling \eref{eq:t-k} becomes 
\numparts
\begin{eqnarray}
  \label{eq:integral-v}
\fl t_{\ve{q}_a, \ve{q}_b} 
   =  t_0 \int_{-2\pi}^{2\pi} dv_1 \int_{-2L_b}^{2L_a} dv_2 \; \frac{e^{-(D(v_1,
        v_2) - \Delta)/a_t}}{A_{\mathrm{cell}}^2\sqrt{4N_aN_b}}
     e^{i v_1(q_{\perp b}R_b + q_{\perp a}R_a)} e^{i v_2(q_{\parallel b} +
      q_{\parallel a})} \\ 
\fl  \times \int_0^{4\pi} du_1 \int_{0}^{2(L_a+L_b)} du_2 \; e^{iu_1(q_{\perp a}R_a  - q_{\perp b}R_b)} e^{i u_2(q_{\parallel a} - q_{\parallel b})} =: A(\ve{q}_a,\ve{q}_b)\; I(\ve{q}_a,\ve{q}_b), \label{eq:integral-u} 
\end{eqnarray}
\endnumparts
with $v_1 = (x_{\perp a}/R_a - x_{\perp b}/R_b)$, $v_2 = x_{\parallel a} - x_{\parallel b}$, and 
$ u_1 = (x_{\perp a}/R_a + x_{\perp b}/R_b)$ and $u_2 = (x_{\parallel a} + x_{\parallel b})$.
We denoted with $A(\ve{q}_a,\ve{q}_b)$ the amplitude of the coupling (it includes all numerical factors), while $I(\ve{q}_a,\ve{q}_b)$ contains the functions which determine the selection rules discussed below and which appear upon performing the integration in \eref{eq:integral-u}.
It reads
\begin{equation}
\label{eq:coupling-t}
I(\ve{q}_a,\ve{q}_b) = 
\tilde{\delta}\left(\pi(q_{\perp a}R_a - q_{\perp b}R_b)\right) 
  \times 
\tilde{\delta}\left(\frac{L_a+L_b}{2}(q_{\parallel a} - q_{\parallel b})\right),
\end{equation}
where $\tilde{\delta}(x) := \sin(x)/x$ and $\ve{q}_a = \ve{k}_a + \ve{G}_a, \ve{q}_b = \ve{k}_b + \ve{G}_b$. The resulting selection rules act differently on the angular and longitudinal degrees of freedom. It is when considering the latter that the issue of incommensurability arises.\\
The additional integration over $v_1,v_2$ yields the amplitude of the coupling $A(\ve{q}_a,\ve{q}_b)$. 
Although the integrals in \eref{eq:integral-v} are finite, the support of the integrand is well within the integration limits, which can therefore be extended to $(-\infty,\infty)$. Thus we find
\begin{eqnarray}
\label{eq:coupling-strength}
\fl A(\ve{q}_a,\ve{q}_b) & = t_k \exp\left\{ -\frac{\Delta a_t}{8 R_a R_b}
\left( q_{\perp a}R_a + q_{\perp b}R_b\right)^2\right\} 
& \times \exp\left\{ -\frac{\Delta a_t}{8} (q_{\parallel a} + q_{\parallel b})^2 \right\}, 
\end{eqnarray}
where $t_k \sim -0.78$eV contains both $t_0$ and all other numerical factors arising from the integrations. From \eref{eq:effective-coupling} it follows that \eref{eq:coupling-strength} 
and \eref{eq:coupling-t} have to be evaluated for $\ve{q}_a = \ve{k}_a + \ve{G}_a,\; \ve{q}_b = \ve{k}_b + \ve{G}_b$, with $\ve{k}_\beta$ satisfying the boundary conditions \eref{eq:bc}.
It clearly shows that contributions from distant regions of the momentum space are exponentially suppressed. For $(k+G) > 2\pi/a_0$ they are already negligible, therefore the sum in \eref{eq:effective-coupling} can be limited to only a few terms.
\subsection{Selection rules}
\label{sec:coupling-selection}
The selection function $I(\ve{k}_a+\ve{G}_a,\ve{k}_b+\ve{G}_b)$ determines whether the coupling between $\ve{k}_a$ and $\ve{k}_b$ is allowed. Note that all integer values of $x/\pi$ are zeroes of $\tilde{\delta}=\sin(x)/x$, except $x = 0$ where $\tilde{\delta}(0) = 1$.\\
{\em Transverse degree of freedom.} The angular momentum $l_{\perp \beta} = k_{\perp\beta}R_\beta$ can take only integer values and $G_{\perp\beta} R_\beta = n_{1\beta} m_{1\beta} + n_{2\beta} m_{2\beta} \in \mathbb{Z}$.
Therefore $\tilde{\delta}$ acts for the transverse degree of freedom $q_\perp R$ in the same way as a normal Dirac $\delta$. \\
{\em Longitudinal degree of freedom.} 
The longitudinal momentum is $k_{\parallel\beta} = \frac{2\pi}{2L_\beta}l_{\parallel\beta}$. The longitudinal component of a reciprocal lattice vector is 
\begin{equation}
\label{eq:g-parallel}
\fl (\ve{G}_{\beta})_\parallel=(n_{1\beta}\ve{b}_{1\beta} + n_{2\beta}\ve{b}_{2\beta})_\parallel
= M_\beta \frac{2\pi}{L_\beta}\left[ (2m_{1\beta}+m_{2\beta}) n_{2\beta} - (m_{1\beta} + 2m_{2\beta})n_{1\beta})\right],
\end{equation} 
where $M_\beta$ is the number of unit cells in shell $\beta$, $L_\beta = M_\beta |\ve{T}_\beta|$.
The value of $q_{\parallel\beta}$ can therefore always be represented as $\frac{\pi}{L_\beta}l'_\beta$.
If both shells are of {\em equal lengths} $L_a = L_b = L$, which is only possible in commensurate DWNTs,
\begin{equation}
\label{eq:select-longitudinal}
\tilde{\delta}\left(\frac{L_a+L_b}{2}(q_{\parallel a} - q_{\parallel b})\right) =
\tilde{\delta}(\pi(l'_{\parallel a} - l'_{\parallel b})) = \delta(l'_{\parallel a} - l'_{\parallel b}).
\end{equation}
In incommensurate DWNTs the two shells always have {\em different lengths} and the proper selection function is $\tilde{\delta}$. However, as \eref{eq:coupling-strength} shows, the amplitude of the coupling decreases strongly with 
the length of $\ve{q}_a,\ve{q}_b$, therefore only a finite region of the reciprocal space is active in the coupling, i.e. gives a non-vanishing contribution. If the mismatch between shell lengths is small enough, we can still approximate $\tilde{\delta}$ by the Dirac $\delta$ in the whole active region. Depending on which level of precision in this approximation we find acceptable, the maximum allowed mismatch can be large or small. We study only DWNTs for which $(L_a+L_b)(q_{\parallel a} - q_{\parallel b})/2 < \pi/4$ in the active region and use the
Dirac-$\delta$ selection rules below.
\numparts
\begin{eqnarray}
  \label{eq:selection-rule}
\label{eq:v-d}
l_{\perp a} + (m_{1 a} n_{1a} + m_{2 a} n_{2a}) &=
l_{\perp b} + (m_{1b} n_{1b} + m_{2 b} n_{2b}), \\
l_{\parallel a} + \mathcal{F}_a(n_{1a}, n_{2a}) &= l_{\parallel b} + \mathcal{F}_b(n_{1b}, n_{2b}),
\label{eq:u-d}
\end{eqnarray}
\endnumparts
with 
\begin{equation*}
\fl \mathcal{F}_\beta (n_{1\beta}, n_{2\beta}) = 2M_\beta \left((2m_{1\beta} + m_{2\beta}) n_{1\beta} - (m_{1\beta} + 2m_{2\beta}) n_{2\beta}\right).
\end{equation*}
In the infinite DWNTs ~\cite{wang:prl2005} the integral over $u_2$ in \eref{eq:integral-u} runs over infinity and the longitudinal selection rule becomes a true Dirac $\delta(q_{\parallel a} - q_{\parallel b})$. The longitudinal momentum is continuous, which means that for any pair of momenta $k_{\parallel a}, k_{\parallel b}$ exists at least one pair of $\ve{G}_a,\ve{G}_b$ such that $q_{\parallel a}, q_{\parallel b}$ fulfill the selection rule, therefore in principle all longitudinal momentum states are coupled. However, in actual computation only the contributions from the active region of the reciprocal space count and many of the couplings vanish, reestablishing the division of the momentum space into independent subspaces, as was the case in finite DWNTs.

\subsection{Angular momentum: coupling between subbands}
\label{sec:coupling-transverse}
\begin{figure}[h]
\begin{center}
 \includegraphics[width=11cm]{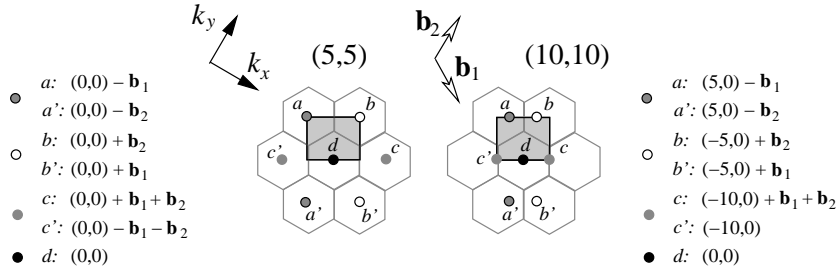}
\end{center}
\caption{Coupled subbands in a (5,5)@(10,10) DWNT. As an example we consider the point $d= (l_{\perp a},l_{\parallel a}) = (0,0)_a$ belonging to the reciprocal space of the inner shell. 
 $\ve{b}_1,\ve{b}_2$ are the graphene reciprocal lattice generators \eref{eq:reciprocal-generators}. After appropriate translations by graphene reciprocal lattice generators $\ve{b}_1,\ve{b}_2$ \eref{eq:reciprocal-generators} the state $(l_{\perp a},l_{\parallel a}) = (0,0)_a$ generates the inner shell states $a,a',b,b'$ and $c,c'$. According to \eref{eq:coupled-subbands} each of these states couples to its counterpart $a,...,c'$ in the outer (10,10) shell, resulting in a non-zero coupling.}
 \label{fig:example-coupling}
\end{figure}

In this section we analyze the implications of \eref{eq:v-d}, i.e. we find the set of states which fulfill the angular momentum selection rules. As an example we consider the commensurate DWNT (5,5)@(10,10). The amplitude function $A$ \eref{eq:coupling-strength} is dominated by the regions in the reciprocal space corresponding to small values of $\ve{G}_a, \ve{G}_b$. Among the reciprocal cells of dominant contributions we find 
\begin{eqnarray} 
\fl i) \hspace{.5cm} \ve{G}_a = (n_{1a}, n_{2a}) = (0,0),\; \ve{G}_b = (0,0) \; &\Rightarrow\;
 	 l_{\perp b} = l_{\perp a}  \nonumber\\
\fl ii) \hspace{.5cm} \ve{G}_a = (-1,0)\; \rm{or}\; (0,-1),\; \ve{G}_b = (-1,0)\; \rm{or}\; (0,-1)  \;&\Rightarrow\; 
  	l_{\perp b} = l_{\perp a} + 5  \label{eq:coupled-subbands}\\
\fl ii) \hspace{.5cm} \ve{G}_a = (1,0)\;\rm{or}\;(0,1), \ve{G}_b = (1,0)\; \rm{or}\; (0,1)  \;&\Rightarrow\; 
  	l_{\perp b} = l_{\perp a} - 5  \nonumber \\
\fl iv) \hspace{.5cm} \ve{G}_a = (1,1)\; \rm{or}\; (-1,-1),\; \ve{G}_b = (1,1)\; \rm{or}\;(-1,-1)  \;&\Rightarrow\; 
  	l_{\perp b} = l_{\perp a} - 10. \nonumber
\end{eqnarray}
For example, $l_{\perp a}=0$ yields in case $i)$ $l_{\perp b} =0$, in case $ii)$ $l_{\perp b} =5$, in case $iii)$ $l_{\perp b}=-5$ and in case $iv)$ $l_{\perp b}=-10$. These coupled states are shown in \fref{fig:example-coupling}, where for clarity only the states with $l_{\parallel a}=0, l_{\parallel b}=0$ are shown.Other combinations of $\ve{G}_a,\ve{G}_b$ would in this DWNT yield the same results. All combinations listed above fulfill also the second selection rule \eref{eq:u-d} for $l_{\parallel a} = l_{\parallel b}$. \\
When we apply the selection rules \eref{eq:v-d} in turn to all the found $l_{\perp b}$, we find other $l_{\perp a}$ states which also couple to the $l_{\perp b}$'s found above. It turns out that in this particular DWNT the sets of coupled states contain only a few elements. 
For each initial $l_{\perp a} = l_0\in [0,4]$, the set of coupled angular momentum values consists of $l_{\perp a} = l_0 -5,l_0$ and $l_{\perp b}=l_0-10,l_0-5,l_0,l_0+5$.  This is a rather unusual situation, occuring only when the chiral indices of one shell are integer multiples of those in the other. In an average DWNT the coupled sets are larger.\\
\subsection{Longitudinal momentum -- the issue of commensurability}
\label{sec:coupling-longitudinal}
\begin{figure}[htbp]
\begin{center}
\includegraphics[width=3cm,angle=-90]{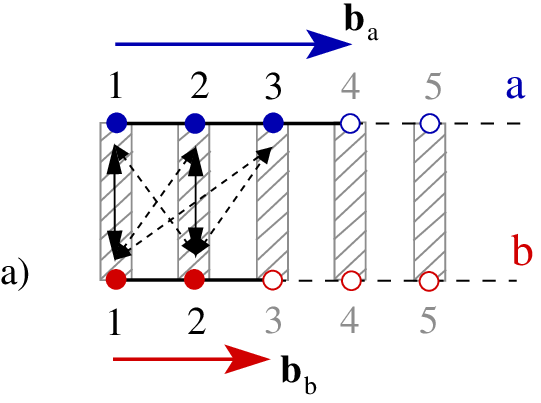}\hspace{2cm}
\includegraphics[width=3cm,angle=-90]{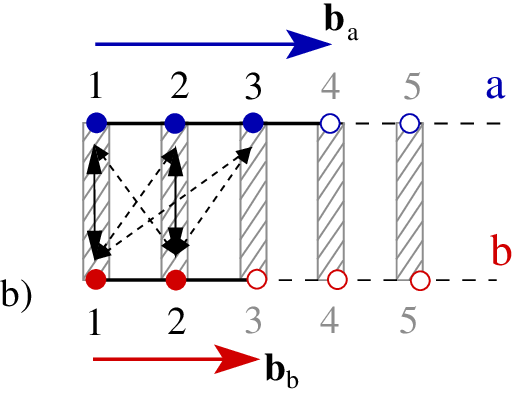}
\end{center}
\caption{Possible couplings between momentum states in commensurate and incommensurate finite chains. Solid black lines mark the first Brillouin zones, $\mathbf{b}_{a}$ and $\mathbf{b}_{b}$ are the reciprocal lattice generators on chain $a$ and $b$, respectively. Filled dots stand for states in the first Brillouin zones, open dots for the states in the rest of the reciprocal space. Dashed grey rectangles mark the regions in which the momenta on different chains match under $\tilde{\delta}$. Black dashed arrows connect states which are coupled after the translational equivalence has been taken into account.
{\bf a)} {\em Commensurate chains with unit cell ratio 2/3. } The couplings between states $1_a,1_b$ and $2_a,2_b$ are the result of direct matching of momenta. The state $3_a$ has the same momentum as $3_b$, but $3_b$ is equivalent to $1_b$ under the translation by $\mathbf{b}_{b}$, which means that $3_a, 1_b$ are also coupled. Similar situation occurs for $4_a$ and $4_b$, which are equivalent to $1_a$ and $2_b$, and so forth. In the end, all states are coupled, although the coupling may be weak. It is possible to define a common Brillouin zone, with the length $2\pi/3a_a = 2\pi/2a_b$. 
{\bf b)} {\em Incommensurate chains with unit cell ratio $1/\sqrt{3}$. } The lengths of the chains 
cannot match - here they are chosen as 3 unit cells of chain $a$ and 2 unit cells of chain $b$. The momenta in chain $a$ are shifted with respect to those in chain $b$, but this mismatch is not large and the same couplings as in the case a) occur. 
 }\label{fig:chains}
\end{figure}

In our calculations we consider only DWNTs in which the two shells have equal or very similar length. If $\Delta L < a_0/8$ we are allowed to use $\delta$-like selection rules and the physical space divides into subspaces containing the coupled longitudinal momentum states. The Hamiltonian matrix acquires a block-diagonal structure with the size of the blocks determined by the geometry of the shells.\\
The active region of the reciprocal space can contain several reciprocal cells contributing to the coupling, which causes the mixing of longitudinal momentum states. The number of involved reciprocal cells and therefore of coupled longitudinal momenta increases with the size of the direct lattice unit cell.\\
If the unit cell ratio of the two shells is rational, $\chi = p/q$ where $p,q\in\mathbb{N}$, the length of the DWNT is $L = qM |\ve{T}_{a}| = pM |\ve{T}_{b}|$. The selection rules split the momentum space into $M$ subspaces, each containing the full set of subbands ($\{ l_{\perp}\}$) for $q$ longitudinal states in shell $a$ and $p$ longitudinal states in shell $b$. The size of each subspace is $2(qS_a + pS_b)$.\\
In the case of incommensurate DWNTs, the lengths of the shells can be chosen so as to minimise $\Delta L$ and allow us to use the exact conservation of crystal momentum, as explained above. If these optimal values of shell lengths can be expressed as $L_a = \tilde{q}M |\ve{T}_{a}|$ and $L_b = \tilde{p}M |\ve{T}_{b}|$, where $\tilde{p},\tilde{q}\in\mathbb{N}$, the Hamiltonian splits into $M$ diagonal blocks, each of the size $2(\tilde{q}S_a + \tilde{p}S_b)$. The ratio $\tilde{p}/\tilde{q}$ is in fact a rational approximation of the irrational $\chi$ and the precision of this approximation depends on the required value of the difference between shell lengths.\\
The difference between the commensurate and incommensurate shells is shown in \fref{fig:chains} in the case of two finite linear chains.
\subsection{Energy spectrum at the Fermi level}
\label{sec:coupling-fermi}
\begin{figure}[h]
\begin{center} 
\includegraphics{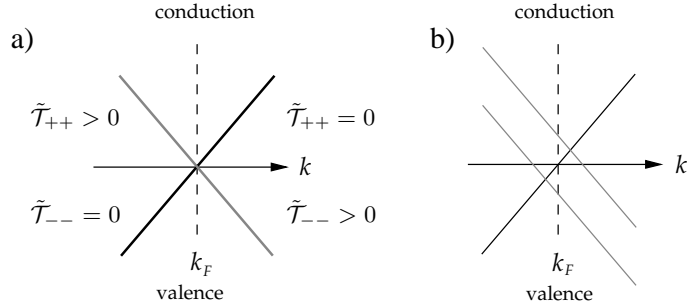}
\end{center}
\begin{picture}(0,0)
\put(23,30){{\small $\tilde{\mathcal{T}}_{++} > 0 $}}
\put(57,30){{\small $\tilde{\mathcal{T}}_{++} = 0 $}}
\put(57,17){{\small $\tilde{\mathcal{T}}_{--} > 0 $}}
\put(23,17){{\small $\tilde{\mathcal{T}}_{--} = 0 $}}
\end{picture}
\caption{Schematic plot of the band crossing at the Fermi level in a (5,5)@(10,10) DWNT. a) The coupling between subbands at $k<k_f$ and $k>k_F$. Only the conduction band is affected in the former, only the valence band in the latter case. b) Resulting shift of the energy levels close to the Fermi point. The parts with negative slope are split, the parts with positive slope remain degenerate.\label{fig:schematic-anticrossing}
}
\end{figure}

The details of the spectrum at the Fermi level depend on the form of both intra- and intershell interaction, most notably on whether the curvature of the nanotube is taken into account or not. Among the effects of curvature in single-wall nanotubes are the rehybridization of $\sigma$ and $\pi$ bonds and varying angle between $\pi$ orbitals \cite{kleiner:prb2001, ding:jpcm2003}. They result in variations in the bond length and bond angle between the lattice atoms, which can cause the opening of a diameter-dependent gap at the Fermi level in metallic SWNTs.
Moreover, the band structure at the Fermi level depends on the relative position of the shells, as it was found to be the case in a (5,5)@(10,10) DWNT studied in \cite{kwon:prb1998,lambin:prb2000}. When the (5,5)@10,10) DWNT is
in a configuration of maximum symmetry $D_{5h}$ \cite{lin:physicae2006, mayer:carbon2004}, the only effect of the intershell hopping is a uniform split and shift of the Fermi subbands, resulting in the presence of four subband crossings. If the symmetry of the system is lowered, four pseudogaps (the largest of the order of 0.1 eV) open in the spectrum \cite{kwon:prb1998}.\\ 
When, as in this work, curvature effects are neglected, only a subband shift is observed (see \fref{fig:anticrossing}) -- in other words, our nanotube is always in the configuration of maximum symmetry. Due to the small size of the curvature-induced gap, we think that our model still yields a reliable description also of band features near the Fermi energy.\\
The presence of a uniform shift between the subbands of a (5,5)@(10,10) DWNT can be understood by considering just the coupling between the Fermi subbands. 
The general Hamiltonian \eref{eq:H-plane} for $\mathbf{k}_a$ and $\mathbf{k}_b$ becomes a 4x4 matrix if all other couplings are ignored. In the sublattice basis it has the form\\
\begin{equation}
\fl H(\mathbf{k}_a, \mathbf{k}_b) = \left(
\begin{array}{cc cc}
0  & |\gamma_a|e^{i\theta_a} & t_{ab} & t_{ab} e^{i\varphi_{AB}} \\ [.25cm]
|\gamma_a| e^{-i\theta_a} & 0 & t_{ab} e^{i\varphi_{BA}} & t_{ab} e^{i\varphi_{BB}} \\[.25cm]
t_{ab} & t_{ab} e^{-i\varphi_{BA}} & 0 & |\gamma_b| e^{i\theta_b} \\[.25cm]
t_{ab} e^{-i\varphi_{AB}} & t_{ab} e^{-i\varphi_{BB}} & |\gamma_b| e^{-i\theta_b} & 0
\end{array}
\right),
\end{equation}
where $\gamma_\beta(\mathbf{k}_\beta) = \gamma_0 \sum_{j=1}^3 \exp(i\mathbf{k}_\beta\cdot\mathbf{d}_j) =: |\gamma_\beta| \exp(i\theta_\beta)$, $t_{ab}$ is the coupling amplitude between $\mathbf{k}_a, \mathbf{k}_b$ from \eref{eq:coupling-strength} and
$\varphi_{\nu\nu'} = i\mathbf{G}_a\cdot\mathbf{\tau}_{a\nu} - i \mathbf{G}_b\cdot\mathbf{\tau}_{b\nu'}$ is the phase associated with hopping between different sublattices. It is clear that $\varphi_{BB} = \varphi_{AB} + \varphi_{BA}$. When this Hamiltonian is expressed with the help of \eref{eq:basis-change} in the valence/conduction basis, it becomes
\begin{equation}
H(\mathbf{k}_a, \mathbf{k}_b) = \left(
\begin{array}{cc cc}
|\gamma_a| & 0  & \tilde{\mathcal{T}}_{++} &  \tilde{\mathcal{T}}_{+-}\\ [.25cm]
0 & -|\gamma_a| &  \tilde{\mathcal{T}}_{-+} &  \tilde{\mathcal{T}}_{--}\\[.25cm]
 \tilde{\mathcal{T}}^*_{++} &  \tilde{\mathcal{T}}^*_{-+} & |\gamma_b| & 0 \\[.25cm]
 \tilde{\mathcal{T}}^*_{+-} &  \tilde{\mathcal{T}}^*_{--} & 0 & -|\gamma_b| 
\end{array}
\right).
\end{equation}
The elements of the coupling matrix in this basis are
\begin{eqnarray}
\fl
 \tilde{\mathcal{T}}_{++} & = & \frac{t_{ab}}{2}\left(
e^{i(\theta_b - \theta_a)} + e^{i(\theta_b + \varphi_{BA})} + e^{-i(\theta_a - \varphi_{AB})}
+ e^{i(\varphi_{AB} + \varphi_{BA})} 
\right),\\
\fl \tilde{\mathcal{T}}_{+-} & = & \frac{t_{ab}}{2}\left(
-e^{i(\theta_b - \theta_a)} - e^{i(\theta_b + \varphi_{BA})} + e^{-i(\theta_a - \varphi_{AB})}
+ e^{i(\varphi_{AB} + \varphi_{BA})} 
\right),\\
\fl \tilde{\mathcal{T}}_{-+} & = & \frac{t_{ab}}{2}\left(
-e^{i(\theta_b - \theta_a)} + e^{i(\theta_b + \varphi_{BA})} - e^{-i(\theta_a - \varphi_{AB})}
+ e^{i(\varphi_{AB} + \varphi_{BA})} 
\right),\\
\fl \tilde{\mathcal{T}}_{--} & = & \frac{t_{ab}}{2}\left(
e^{i(\theta_b - \theta_a)} - e^{i(\theta_b + \varphi_{BA})} - e^{-i(\theta_a - \varphi_{AB})}
+ e^{i(\varphi_{AB} + \varphi_{BA})} 
\right).
\end{eqnarray}
In order to analyze the nature of the subband (anti)crossing at the Fermi level, we need to evaluate $\gamma(\mathbf{k}_\beta)$, $t_{ab}$, $\varphi_{AB}$ and $\varphi_{BA}$. Let us begin by the in-shell part.\\
For Fermi subbands the angular momentum is set to $l_{\perp a} = -5$ and $l_{\perp b} = -10$. We can rewrite $\gamma(\mathbf{k}_\beta)$ as a function of the distance between longitudinal momentum and the Fermi point, $\Delta k = k_\parallel - k_F$. Both subbands have the same position in the reciprocal cell of the armchair nanotube, therefore $\gamma_a(\Delta k) = \gamma_b(\Delta k)$. The examination of $\gamma(\Delta k)$ reveals that its phase has only two values:
\begin{equation}
\gamma(\Delta k) = \left|\gamma_0 \left(2\cos\left( \frac{\pi}{3}
+ \frac{\sqrt{3}}{2} \Delta k a_0\right)\right)\right| \times \left\{
\begin{array}{ll}
 e^{i2\pi/3}, & \Delta k < 0 \\
e^{-i\pi/3}, & \Delta k > 0
\end{array}
\right. .
\end{equation}
In the inter-shell part we have to perform a sum over reciprocal lattice vectors as in \eref{eq:effective-coupling}. The vectors which give the dominant contribution to the coupling are $\mathbf{G}_a = \mathbf{b}_1, \mathbf{G}_b = \mathbf{b}_1$. The phases associated with hopping between sublattices are then
\begin{equation}
\varphi_{AA} = 0, \hspace{.5cm} \varphi_{AB} = \frac{2\pi}{3}, \hspace{.5cm}\varphi_{BA}=-\frac{2\pi}{3}, \hspace{.5cm}\varphi_{BB} = 0. 
\end{equation}
We have now a situation where $\theta_a = \theta_b = \theta$ and $\varphi_{AB} = -\varphi_{BA} = \varphi$. The coupling matrix $\tilde{\mathcal{T}}$ becomes
\begin{equation}
\fl \tilde{\mathcal{T}} = t_{ab} \left(
\begin{array}{cc}
1 + \cos(\theta-\varphi) & -i \sin(\theta - \varphi) \\
i \sin(\theta - \varphi) & 1 - \cos(\theta - \varphi) 
\end{array}
\right)
=
t_{ab} \left\{
\begin{array}{ll}
 \left(
	\begin{array}{cc}
	2 & 0 \\ 0 & 0
	\end{array}
 \right), & \Delta k < 0 \\[.25cm]
 \left(
	\begin{array}{cc}
	0 & 0 \\ 0 & 2
	\end{array}
 \right), & \Delta k > 0 \\
\end{array}
\right. 
\end{equation}
The coupling does not mix bands and moreover affects only the conduction band for $k<k_F$ and the valence band for $k>k_F$, as shown in \fref{fig:schematic-anticrossing}. In consequence, the negative slope parts of the two subbands $-5_a$ and $-10_b$ are split evenly on both sides of the Fermi point. The results of our calculation show that there is also a smaller uniform split of the part with positive slope (see \fref{fig:anticrossing}), which is due to the smaller couplings between $-5_a$ and $-5_b$ and $5_b$.
This asymmetry is explained in \cite{nemec:phd2007} in real space terms as the result of different phases of the wavefunctions on sublattices $A$ and $B$. The wavefunctions belonging to the negative slope parts of the subbands have constant phase on the whole circumference, while the wavefunctions belonging to the unshifted parts have different phases on sublattices $A$ and $B$, therefore they cannot hybridize so well. 

\subsection{Results}
\label{sec:results-coupling}
\begin{figure}[h]
\begin{center}
\includegraphics[height=4.5cm]{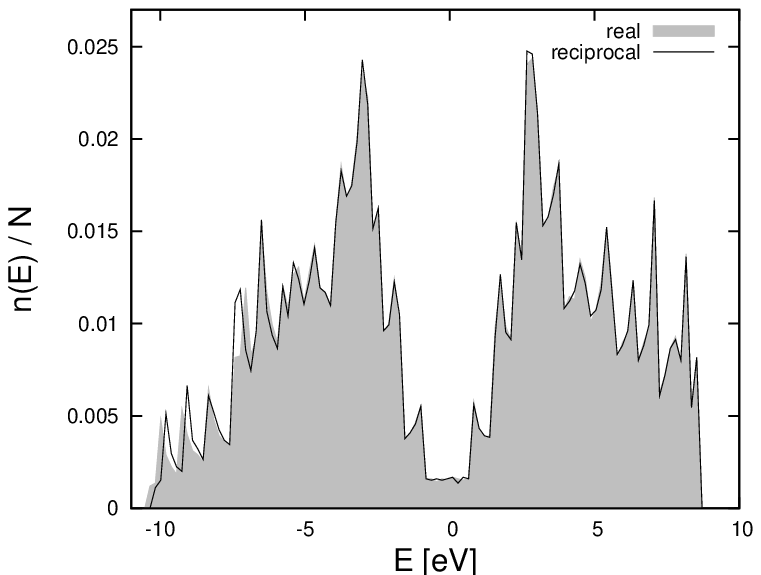}
\includegraphics[height=4.5cm]{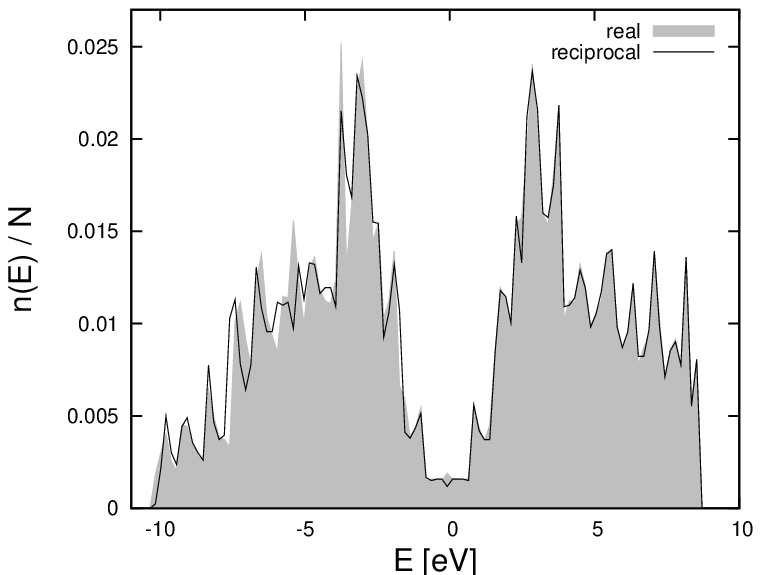}
\end{center}
\caption{Comparison between the DOS of a (5,5)@(10,10) (left panel) and a (9,0)@(10,10) (right panel) DWNT evaluated by diagonalizing the DWNT Hamiltonian in real (Eq. \eref{eq:H-total}) and reciprocal (Eq.  \eref{eq:H-plane}) space. $n(E)/N$ is the density of states normalized to 1 ($N$ is the number of atoms in the nanotube). In (5,5)@(10,10) both shells have 120 unit cells and are 30 nm long. In the (9,0)@(10,10) the zigzag shell contains 75 unit cells, the armchair - 130, corresponding to the DWNT length of approximately 32 nm. Notice the breaking of the electron-hole symmetry due to the intershell tunneling.}
\label{fig-compare-xk}
\end{figure}

\begin{figure}[h]
\begin{center}
\includegraphics{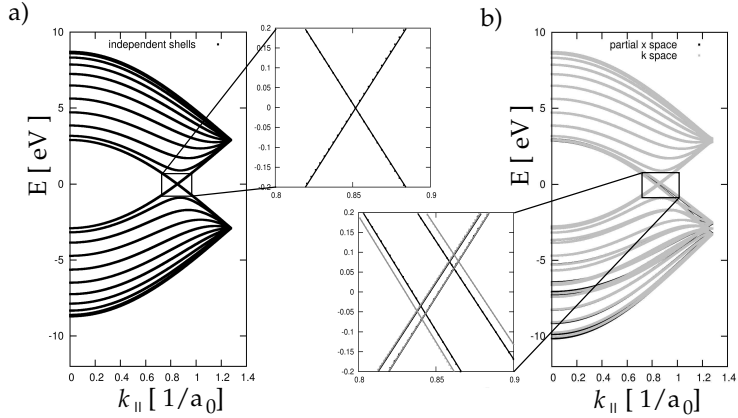} 
\end{center}
\caption{\label{fig:anticrossing}
Anticrossing at the Fermi level in a (5,5)@(10,10) DWNT. For independent shells (a) the subbands at the Fermi level are degenerate, but when the intershell tunneling is allowed (b), the degeneracy is removed. The energy levels in the ``partial real space'' method were obtained by defining a DWNT supercell and using the Bloch theorem in the longitudinal direction. 
}
\end{figure}

In order to test our method we calculated the electronic spectra of a short commensurate (5,5)@(10,10) and a short incommensurate (9,0)@(10,10) DWNT, both by the $k$-space method described above and by direct diagonalization of the Hamiltonian in the real space. Our test DWNTs have both shells of equal or very similar length.
The commensurate DWNT consists of 120 unit cell lengths of both the inner and outer armchair. The incommensurate DWNT has 75 unit cells of the zigzag shell and 130 of the armchair. The mismatch between shell lengths is 0.17$a_0$, which still allows us to use the exact selection rules \eref{eq:u-d}. The spectra calculated by both methods match well (\fref{fig-compare-xk}).\\
The asymmetry between the valence $(E<0)$ and conduction $(E>0)$ bands, seen also in \cite{uryu:prb2004}, is due to the intershell tunneling. The wave functions of the coupled momentum states hybridize and form bonding and antibonding combinations, with greatest energy differences in the bottom of the valence band. \\
The coupling between momentum states is felt most strongly by states with low momentum, but its consequences can also be seen at the Fermi level.
In our model, which neglects curvature effects, the band structure at the Fermi level does not depend on the shell shift. In the k-space approach the relative position of the shells enters only through a phase factor in the coupling matrix \eref{eq:effective-coupling} and does not affect the energy eigenvalues. We have tested our prediction for a (5,5)@(10,10) DWNT using a ``partial real space'' method. We defined a supercell containing one unit cell of the outer and one of the inner shell. Then we used the Bloch theorem in the longitudinal direction and the spectrum which we obtain is also insensitive to the shell shift (\fref{fig:anticrossing}).\\ 

\section{DWNT in parallel magnetic field}
\label{sec:magnetic}

\subsection{The intershell coupling}
\begin{figure}[htpb]
\begin{center}
   \includegraphics[width=4cm,angle=-90]{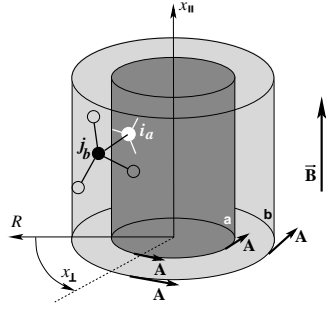}
\end{center}
\caption{DWNT in a uniform magnetic field parallel to its axis in tangential gauge. The atoms $i_a$ on the inner shell $a$ and $j_b$ on the outer shell $b$ bind to their in-shell neighbours and with each other. A phase factor comes into the Hamiltonian with each bond.}
\label{fig:dwnt-magnetic}
\end{figure}

\begin{figure}
\begin{center}
 \includegraphics[height=4.5cm,angle=-90]{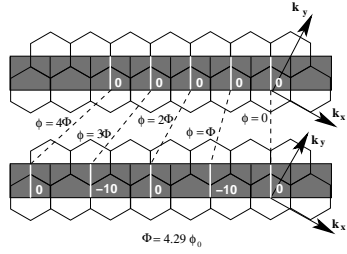}
\end{center}
\caption{Subbands coupled to $l_{\perp a} = 0$ in the inner shell of a (5,5)@(10,10) DWNT and the reciprocal cells of the main contributions. As the magnetic field increases, more distant cells are involved and the dominant coupling switches between $(l_{\perp a},l_{\perp b}) = (0_a,0_b)$ and $(0_a,-10_b)$ with the period $\Phi\approx4.29\phi_0$.}
\label{fig:coupling-shift}
\end{figure}

When a magnetic field is applied to a system, it usually changes the system's symmetries, since the vector
potential $\mathbf{A}$ depends on the spatial coordinates. As a consequence the wave function of a charged particle moving in the magnetic field gathers a phase factor during its motion. This is due to the modification of the momentum operator, $\mathbf{p}\rightarrow\mathbf{p} - q\mathbf{A}$, known as the minimal coupling principle or the Peierls substitution. The translation operator $T$, where $T(\mathbf{x})\psi(\mathbf{r}) = \psi(\mathbf{r} + \mathbf{x})$ ($\ve{r}$ is a position vector, $\ve{x}$ is the translation vector), is modified accordingly by the {\em Peierls phase}
\cite{peierls:zphys1933}:
\begin{equation}\label{eq-translation}
\fl T(\mathbf{x})  = \exp\left(\frac{i}{\hbar}\mathbf{x}\cdot\mathbf{p}\right)
\hspace{0.5cm}\rightarrow\hspace{0.5cm}
T'(\mathbf{x}) = \exp\left\{ \frac{iq}{\hbar} \int_{\mathbf{r}}^{\mathbf{r}
+ \mathbf{x}} \mathbf{A}(\mathbf{r}')\cdot d\mathbf{r}'
\right\}T(\mathbf{x}).
\end{equation}
In a uniform field the Hamiltonian remains invariant under translations.
 For the lattices considered in the tight-binding model it implies that each hopping integral is modified by the appropriate phase factor.\\
 The influence of the magnetic field has been most extensively studied in two simplest cases - of a uniform magnetic field perpendicular to the flat lattice, and a uniform magnetic field parallel to the axis of a system with cylindrical topology. \\
%
%
In the former case the application of the magnetic field changes or destroys the periodicity of a lattice. If the magnetic flux through the area of the elementary cell is rational, $\phi_{cell} = p/q\; \phi_0$, where $p,q \in \mathbb{Z}$ and $\phi_0=e/h$ is the flux quantum, it is possible to define an enlarged elementary cell, containing $q$ original ones, pierced
by $p$ flux quanta. Thus the lattice remains periodic, although with a different period. If the flux through the elementary cell is irrational, $\phi_{cell}/\phi_0 \notin \mathbb{Q}$, the periodicity is removed altogether and the spectrum of the system is fractal. The plots of the energy spectrum vs. $\phi_{cell}$ are known as ``Hofstadter butterflies'' \cite{hofstadter:prb1976}.\\
%
%
In systems such as rings and cylinders, a uniform magnetic field parallel to the axis gives rise to the Aharonov-Bohm effect. In simple systems its consequence is a shift of all the angular momentum states by the number of flux quanta flowing through its cross-section, $\phi_{cross-section}/\phi_0$. The change in 
the spectrum is periodic with a period $\phi_0$ \cite{buttiker:prb1985,gefen:prb1988,ajiki:jpsj1993a}. In DWNTs tunneling can also occur between shells and the Peierls phase enters not only into the in-shell term, but also into the inter-shell hopping (see \fref{fig:dwnt-magnetic}) and \eref{eq:H-total} becomes
\begin{eqnarray}\label{eq-hamil}
\fl \eqalign{H (\mathbf{A}) & =\sum_{\beta\sigma} \sum_{\langle i,j \rangle} \gamma_0
\exp\left\{\frac{ie}{\hbar}
\int_{\mathbf{r}_{\beta j}}^{\mathbf{r}_{\beta i}}\mathbf{A}(\mathbf{r}')\cdot
d\mathbf{r}'\right\}c^\dag_{\beta i\sigma} c_{\beta j\sigma} \cr
\fl & + \left[ \sum_{i,j,\sigma} t(\ve{r}_{a i}, \ve{r}_{b j}) \exp\left\{\frac{ie}{\hbar}
\int_{\mathbf{r}_{b j}}^{\mathbf{r}_{a i}}\mathbf{A}(\mathbf{r}')\cdot
d\mathbf{r}'\right\}c^\dag_{a i\sigma} c_{b j\sigma} + h.c. \right].}
\end{eqnarray}
In the cylindrical coordinates and the tangential gauge we have chosen, $\mathbf{A} = (A_r, A_\varphi, A_z) = (0,  B r/2, 0)$. The phase factors attached to in-shell bonds are
\begin{equation}\label{eq-in-shell}
\fl 
\exp\left\{\frac{ie}{\hbar}
\int_{\mathbf{r}_{\beta i}}^{\mathbf{r}_{\beta j}}\mathbf{A}(\mathbf{r}')\cdot
d\mathbf{r}'\right\} =
e^{i\frac{\phi_\beta}{\phi_0}\left( \varphi_{\beta i} - 
\varphi_{\beta j}\right)} = e^{i\frac{\phi_\beta}{\phi_0}\left( \frac{x_{\perp \beta i}}{R_\beta} - 
\frac{x_{\perp \beta j}}{R_\beta}\right)},
\end{equation}
where $\phi_\beta$ is the magnetic flux through the shell $\beta$. The dispersion relation contains therefore a dependence on the magnetic flux:
\begin{equation}
\label{eq:dispersion-magnetic}
\fl \epsilon_{\beta,\nu}(\ve{k},\phi_\beta) = \epsilon_{\beta,\nu}((k_{\perp\beta},k_{\parallel\beta}),\phi_\beta) =
\epsilon_{\beta,\nu}\left(\left(k_{\perp\beta}+\frac{\phi_\beta}{\phi_0}\frac{2\pi}{C_{h\beta}},k_{\parallel\beta}\right), 0\right).
\end{equation}
The phase factor in the term describing the inter-shell interaction is
\begin{eqnarray}\label{eq-inter-shell}
\fl
\eqalign{
\exp\left\{ \frac{ie}{\hbar}\int^{\mathbf{r}_{b j}}_{\mathbf{r}_{a i}}
\mathbf{A}(\mathbf{r}')\cdot d\mathbf{r}'\right\}  & =
\frac{i}{3}\frac{\phi_a}{\phi_0}\left(\frac{x_{\perp b j}}{R_b} - \frac{x_{\perp a i}}{R_a}\right)
\left( 1 + \frac{R_b}{R_a} + \left(\frac{R_b}{R_a}\right)^2\right) \cr
 & =: i \left(\frac{x_{\perp b j}}{R_b} - \frac{x_{\perp a i}}{R_a}\right) F\left( \frac{\phi_a}{\phi_0}\right).
}
\end{eqnarray}    
The inter-shell coupling is analogous to \eref{eq:effective-coupling} except that $t_{\ve{q}_a,\ve{q}_b}$ depends on the magnetic flux threading the DWNT. It is given by
\numparts
\begin{eqnarray}
\fl 
t_{\mathbf{q}_a,\mathbf{q}_b} (\phi_a) = & t_k\; 
 \tilde{\delta}\left( \pi 
	(q_{\perp b}R_b - q_{\perp a} R_a)
	\right)
\tilde{\delta} \left( \frac{L_a + L_b}{2} 
	(q_{\parallel b} - q_{\parallel a})
	\right)
			\label{eq:coupling-delta}\\
\fl & \times   
\exp\left\{
	-\frac{\Delta a_t}{8}
	\left(
		q_{\parallel b} + q_{\parallel a}
	\right)^2
\right\} 
			\label{eq:coupling-length}\\
\fl & \times 
\exp\left\{
	-\frac{\Delta a_t}{8 R_a R_b} 
	\left( 
		q_{\perp b}R_b + q_{\perp a}R_a
	+ 2 F\left(\frac{\phi_a}{\phi_0}\right)
\right)^2
\right\},  \label{eq:coupling-angle}
\end{eqnarray}
\endnumparts
where $t_k \sim 0.78$eV. The magnetic field $\ve{B}$ enters the Hamiltonian only as the flux through inner and outer shell. The flux through the outer shell can be expressed as $\phi_b = (R_b/R_a)^2\phi_a$ and in the following we shall present all quantities depending on the magnetic field as functions of the flux through the inner shell, $\phi_a$.\\
%
%
As we can see from the $\tilde{\delta}$ function in \eref{eq:coupling-delta}, the selection rules do not depend on the magnetic field. They determine once and for all the quantum numbers of the coupled states, although for some pairs the main contribution to the sum in \eref{eq:coupling-delta} may come from a very distant reciprocal cell.
The strength of the coupling, however, does depend on the amount of magnetic flux through the system. At vanishing field the most strongly coupled states are those with low momentum; at higher
fields the maximum coupling can occur between states with energy close to the Fermi level or even to the top of the conduction band. From \eref{eq:coupling-angle} we see that the strength of the coupling between angular momenta
evolves with the magnetic field, while the coupling between longitudinal states \eref{eq:coupling-length} remains unchanged. It is to be expected, because the longitudinal motion of the electron does not accumulate the Peierls phase. It is clear that as the flux through the DWNT is increased, the dominant terms in the sum \eref{eq:coupling-delta} come from reciprocal cells with varying $G_\perp$ but constant $G_\parallel = 0$. \\
Let us analyse the influence of the magnetic field on the coupling between individual subbands. As an example we take the $l_{\perp a}=0$ and $l_{\perp b} = -10,l_{\perp b} = 0$ subbands of the (5,5)@(10,10) DWNT. The value of $l_{\parallel a} = l_{\parallel b} = 1$ shall be assumed implicitly.
In the absence of the magnetic field the coupling between subbands $(0_a,0_b)$ is dominant, while the coupling between $(0_a,-10_b)$ almost vanishes.
As we increase the magnetic field, the predominant coupling 
switches between $(0_a,0_b)$ and $(0_a,-10_b)$, while also oscillating in amplitude. The switching occurs 
periodically and the period can be evaluated from \eref{eq:coupling-angle}. The maxima of
the coupling occur when the exponent vanishes:
\begin{equation}
\fl (l_{\perp a} + n_a S_a) + (l_{\perp b} + n_b S_b) + \frac{2}{3}\left(1 + \frac{R_b}{R_a} + \left(\frac{R_b}{R_a}\right)^2\right)\frac{\phi_{a\,max}}{\phi_0} = 0.
\end{equation}
The first maximum of $(0_a,0_b)$ coupling occurs at $\phi_a = 0$. The next maximum coupling is between $(0_a,-10_b)$ and occurs at
\begin{equation*}
 -10 - 10 + 4.67\frac{\phi_{a\,max}}{\phi_0} \hspace{.5cm}\Rightarrow \phi_{a\,max} \approx 4.29 \phi_0.
\end{equation*}
The period of the oscillation of the coupling amplitude is $\Phi \approx 4.29\phi_0$. The switching between
dominant couplings ($(0_a,0_b)$ and $(0_a,-10_b)$), depending on which reciprocal cell is active, is shown in \fref{fig:coupling-shift}. 
With the magnetic field increasing from 0, the reciprocal cells of the dominant contribution change in a sequence
\begin{eqnarray*}
\phi_a = 0, & \ve{G}_a = 0 ,\hspace{.25cm}\ve{G}_b = 0 \\
\phi_a = \Phi, & \ve{G}_a = -1 (\ve{b}_1+\ve{b}_2),\hspace{.25cm}\ve{G}_b = 0 \\
... \\
\phi_a = 2n\Phi, & \ve{G}_a = -2n (\ve{b}_1+\ve{b}_2),\hspace{.25cm}\ve{G}_b = -n (\ve{b}_1 + \ve{b}_2)\\
\phi_a = (2n+1)\Phi,\; & \ve{G}_a = -(2n+1) (\ve{b}_1+\ve{b}_2),\hspace{.25cm}\ve{G}_b = -n (\ve{b}_1 + \ve{b}_2),
\end{eqnarray*}
where $n=1,2...$. If the origins of the shells are aligned, $\ve{\rho}_a = \ve{\rho}_b=0$ (see \fref{fig:dwnt}), the phase factors from \eref{eq:effective-coupling} change with the period $6\Phi$, common to all pairs of coupled states. If any of the $\ve{\rho}$'s is non-zero, the factor $\exp(i \ve{G}\cdot\ve{\rho})$ is periodic in $\phi$ only for $\rho=q a_0$, with $q$ rational. Otherwise the phase factors in \eref{eq:effective-coupling} vary in the magnetic field without showing any periodicity.\\
In DWNTs where the chiral indices of the outer shell are not integer multiples of those in the inner shell, the coupling between $l_{\perp a}=0$ and its partners in the outer shell also strengthens and weakens periodically. For example, in a (6,6)@(11,11) armchair nanotube the subband $0_a$ is coupled to all the even-numbered subbands in the outer shell. In the absence of the magnetic field the dominant coupled pair is $(0_a,0_b)$. As the field increases, the dominant pair becomes $(0_a,10_b)$, then $(0_a,-2_b)$, and so forth. The distance between subsequent maxima of the coupling strength can be found as above, by minimizing the exponent in \eref{eq:coupling-angle} and in this particular DWNT it is $\Phi \approx 5.81\phi_0$. In an incommensurate (9,0)@(10,10) the situation is analogous and the oscillation period is $8.16\phi_0$. 

\subsection{Results}
\begin{figure}
\begin{center}
 \includegraphics[width=12cm,angle=-90]{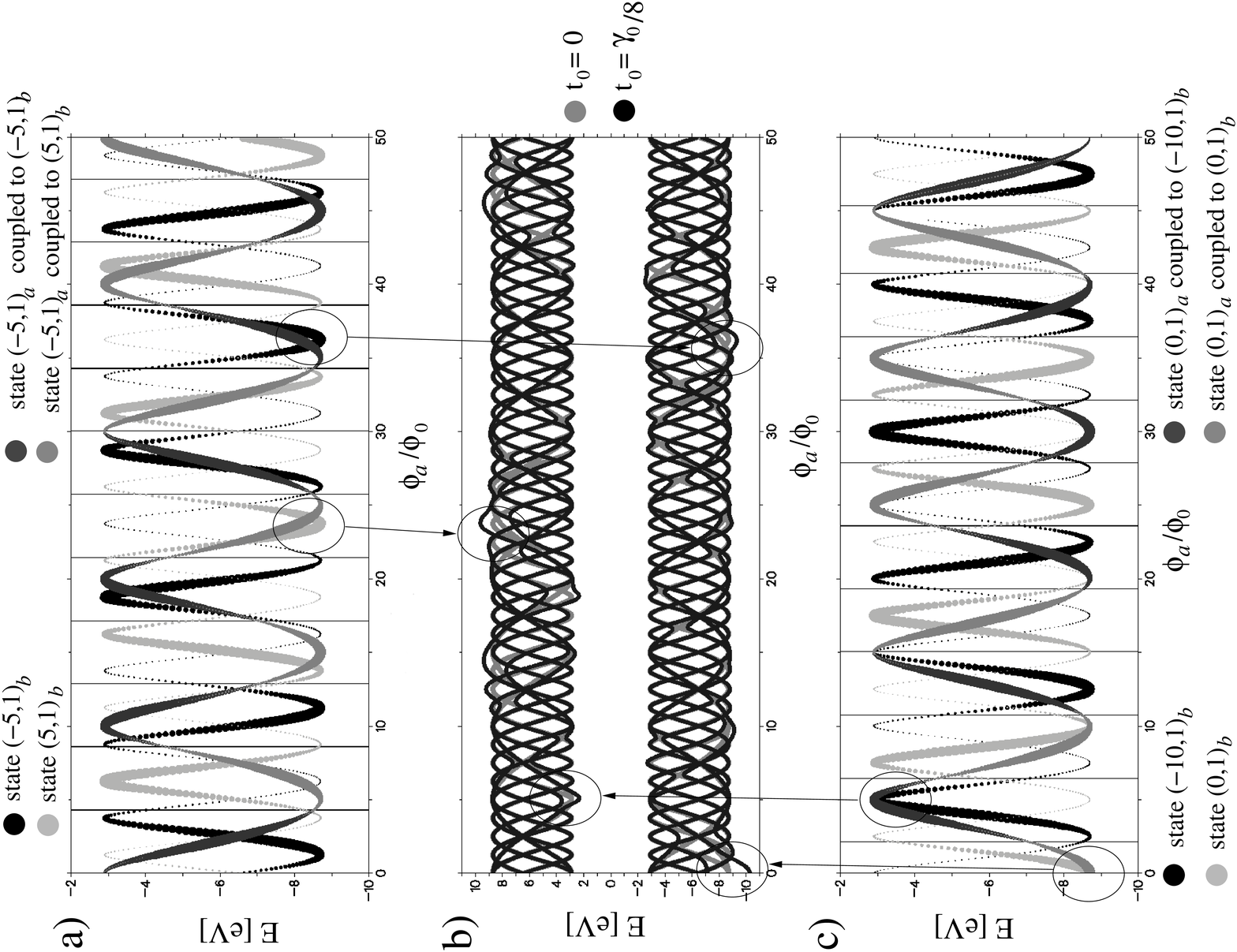}
\end{center}
\caption{The evolution of two sets of coupled states in a (5,5)@(10,10) DWNT. The inner shell states are $(l_{\perp a},l_{\parallel a})\in\left\{(-5,1),(0,1) \right\}$, the outer shell $(l_{\perp b}, l_{\parallel b})\in\left\{(-10,1),(-5,1),(0,1),(5,1) \right\}$. The abscissa corresponds to the inner flux in $\phi_0$ units, the ordinate to the energy. {\bf a)} The energy of the states $(-5,1)_b$ (black), $(5,1)_b$ (light grey) and $(-5,1)_a$. The latter is shown either in dark grey, if the prevailing coupling is $(-5,1)_a$ with $(-5,1)_b$ or in medium grey if $(-5,1)_a$ with $(5,1)_b$ dominates. The width of the lines gives additional information about the size of the coupling, e.g. at $\phi_a \approx 24\phi_0$ the prevailing coupling is between $(-5,1)_a$ and $(5,1)_b$ (wide medium grey and light grey lines), which are also close in energy. The state $(-5,1)_b$ has a much higher energy and couples to $(-5,1)_a$ very weakly (small black dots). At border values of $\phi_a$ where the dominant coupling switches phase, black lines are drawn. {\bf b)} The difference between the energy spectrum for this subspace obtained without inter-shell tunneling ($t_0 = 0$) and with the tunneling of the magnitude $t_0 = \gamma_0/8$.
{\bf c)} The analogon of a), but for the sets of states $(-10,1)_b$ (black), $(0,1)_b$ (light grey) and $(0,1)_a$ (medium or dark grey, depending on the prevalent coupling). It can be seen that at flux values where the energies of strongly coupled states are close, there is a distinctive avoided crossing in the energy spectrum, such as e.g. at $\phi_a=0$ or $\phi_a \approx 4\phi_0$ in c) or $\phi_a \approx 24\phi_0$ and $\phi_a \approx 36\phi_0$ in a). The corresponding regions of large avoided crossings are marked. Whether the crossing occurs in valence or conduction band is determined by the phase factor \eref{eq:t-k} for the Brillouin zone of the dominant contribution. 
}
\label{fig:full-evolution}
\end{figure}

\begin{figure}[h]
\begin{center}
\includegraphics[height=5cm]{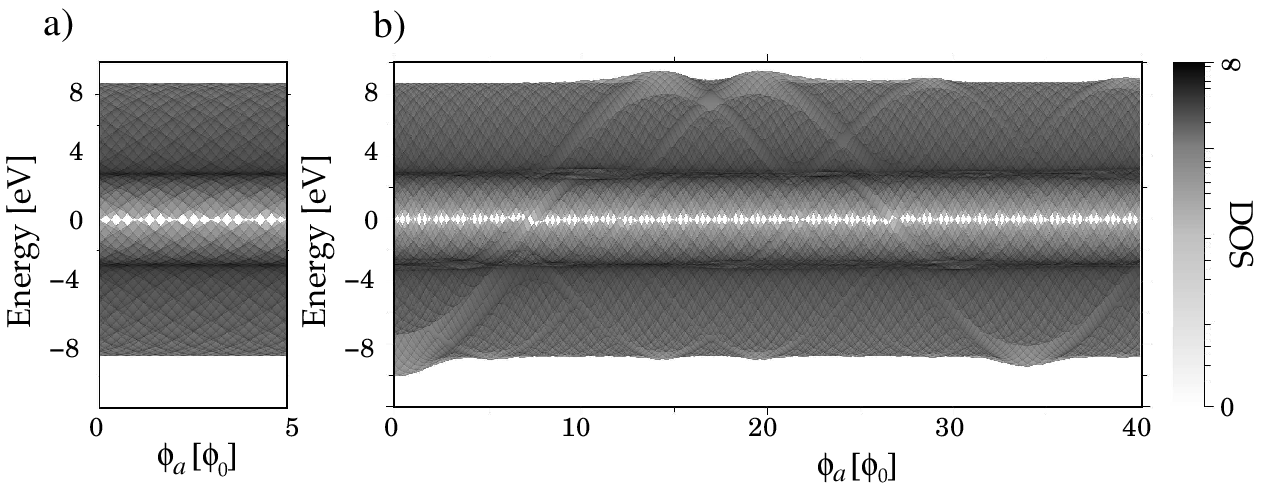}
\end{center}
\caption{The density of states in a (6,6)@(10,10) DWNT in changing parallel magnetic field. a) Without inter-shell tunneling the DOS is a sum of the DOS in both shells. b) The inter-shell coupling causes a change in the DOS varying with the strength of the magnetic field. Here the coupling constant is $t_0=\gamma/8$.
}\label{fig:dos-aina}
\end{figure}

\begin{figure}[h]
\begin{center}
\includegraphics[height=5.5cm]{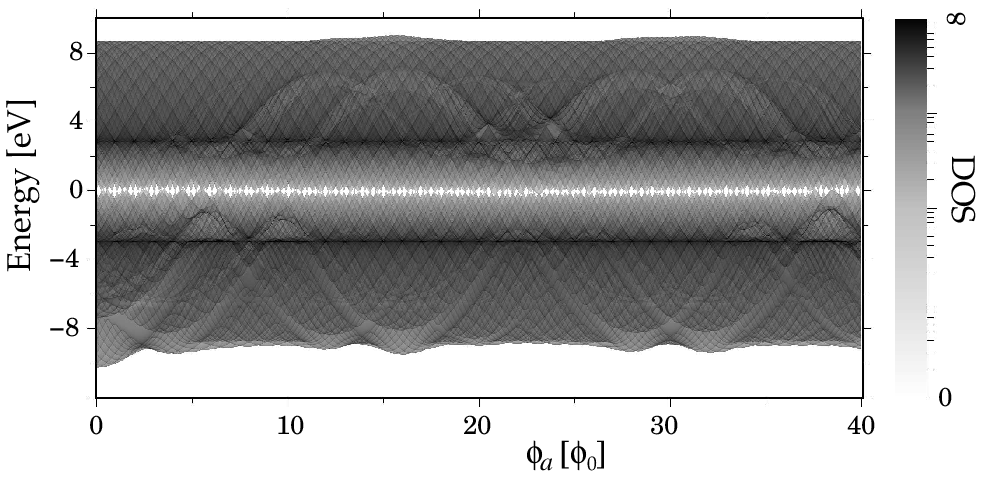}
\end{center}
\caption{The DOS of an incommensurate (9,0)@(10,10) DWNT in changing magnetic field.
Note that in most of the energy range the density of states  seems decreased in comparison with the commensurate case in Fig. \ref{fig:dos-aina}. That effect is due to a very high DOS at the van Hove singularities occurring at $\pm\gamma$, typical for the zigzag nanotubes.
}\label{fig:dos-incom}
\end{figure}

We performed numerical calculations of the DWNT spectra and explored the evolution of coupled states
for several combinations of chiralities. In \fref{fig:full-evolution} we show the
behaviour of the coupling between two sets of states of a (5,5)@(10,10) in the magnetic field. Those sets are $(l_{a\perp},l_{a\parallel})\in\left\{(-5,1),(0,1)\right\}$ in the inner shell and $(l_{b\perp},l_{b\parallel})\in\left\{(-10,1),(-5,1),(0,1),(5,1) \right\}$ in the outer.
In the regions where the coupled states have both similar energies and strong coupling we notice the appearance of avoided crossings. Their size and position in the spectrum is governed by four factors depending on the magnetic field:\\
$i)$ the amplitude of the coupling \eref{eq:coupling-length},\eref{eq:coupling-angle}\\
$ii)$ the phase factors \eref{eq:effective-coupling}\\
$iii)$ the dispersion relations $\epsilon_{a,\nu}(\ve{k},\phi_a/\phi_0)$\\
$iv)$ and $\epsilon_{b,\nu}(\ve{k},\phi_a/\phi_0)$ \eref{eq:dispersion-magnetic}.\\
In this particular case the spectrum is periodic in $\phi_a$ -- the strength of the coupling oscillates with the period $\Phi = 30\phi_0/7$, the phase factors with the period $6\Phi$, the energy of the inner states with the period $\phi_0$, the energy of the outer states with the period $(R_b/R_a)^2\phi_0 = 4\phi_0$. At $\phi_a = 180\phi_0$ the initial spectrum is recovered. Nevertheless, in other DWNTs $R_b/R_a$ is usually irrational and it is in general impossible to find a common period for these four functions.\\
The numerically calculated DOS plots of several nanotubes show features absent in uncoupled DWNTs (see  \fref{fig:dos-aina}). The coupling between states from the two shells causes a series of
avoided crossings, resulting in a whole region in which the density of states is depleted, observed also in ~\cite{nemec:prb2006}. In small magnetic field this region is at the bottom of the valence band, where the momenta
in both shells are small. As the inter-shell tunneling evolves with the increasing magnetic field and higher momentum states become involved, the main region with avoided crossings shifts also towards 
higher energies. \\
These snake-like patterns are a statistical result, caused by many states. Their details vary according to the chiralities of the DWNT's shells. In armchair nanotubes the most strongly coupled states are at the band edges, and the avoided crossing affects van Hove singularities. In other nanotubes the
strongest coupling can occur farther from the band edges, especially when the magnetic field is large. These energy gaps evolving in the middle of the band give rise to less distinctive snake-structures
which can be seen seen in \fref{fig:dos-incom}, where they are caused by the presence of a zigzag shell. 

\begin{figure}[h]
 \begin{center}
  \includegraphics{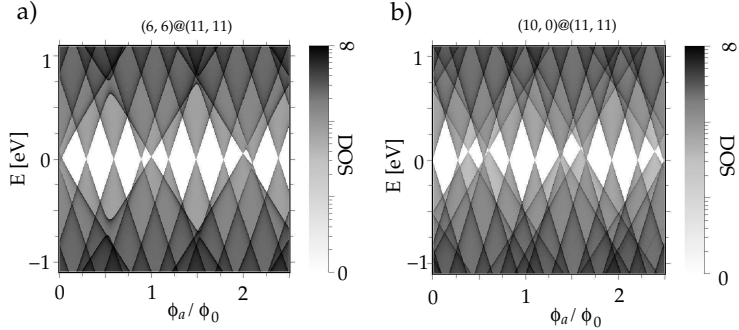}
 \end{center}
\caption{\label{fig:highRes-zoom} The evolution of the DOS at the Fermi level with magnetic field in a commensurate (6,6)@(11,11) (a) and incommensurate (10,0)@(11,11) (b) DWNT. The closing and opening of the gap can be seen in both cases. In (a) the peaks belonging to different shells are mixed, in (b) the only effect of the intershell coupling is a shift of the band crossing in the zigzag shell.}
\end{figure}

The evolution of the density of states with the magnetic field is a superposition of two patterns at different scales. Features with steep $E/\phi$ slope are caused by the outer shell, which feels a flux greater than that in the inner shell and evolves faster with the magnetic field. The features with mild slope are due to the inner shell. \\
Characteristic of the evolution of the DOS near the Fermi level with the magnetic field is the periodic opening and closing of the gap, causing a series of metal-semiconductor transitions, predicted in \cite{ajiki:jpsj1993b} and observed a few years ago in \cite{mceuen:nature2004,coskun:science2004}. These can be seen as the empty diamonds along the $E=0$ line in \fref{fig:highRes-zoom}. In a (6,6)@(11,11) nanotube the intershell tunneling mixes the subbands from both shells, increasing the DOS at the Fermi level whenever both nanotubes have closed gaps, e.g. at $\phi = 0$ and $\phi \approx 2\phi_0$ in \fref{fig:highRes-zoom}a). In an incommensurate (10,0)@(11,11) DWNT the coupling between shells affects at the Fermi level only the DOS of the inner shell -- due to the coupling the band crossing is shifted towards positive energies -- the large diamonds are shifted with respect to the small ones in \fref{fig:highRes-zoom}b). This is a consequence of the structure of the reciprocal cell of the zigzag shell, where the Fermi point (subbands crossing) is at $l_\parallel = 0$, while for the armchair it is at $l_\parallel = 2l_F/3$. The amplitude of the coupling decreases exponentially with the value of the momentum, therefore the zigzag subbands at the Fermi level are affected more strongly by the coupling than the armchair subbands. In transport experiments, where mostly the outer shell is probed, this shift might be visible if the zigzag shell is on the outside.

\section{Conclusions}
\label{sec:results}
In this work we started from an inter-shell tunneling Hamiltonian given in the real space and derived its equivalent in the reciprocal space. In a commensurate (5,5)@(10,10) DWNT the band structure obtained with this method agrees with that obtained by the partial real-space method described in \sref{sec:results-coupling}, down to the fine details of the subband crossings near the Fermi level. As shown in \sref{sec:coupling-fermi}, this method allows us also to study the spectrum near the Fermi level analytically. Although for small nanotubes the curvature (which we neglect) can cause a dependence of the spectrum at the Fermi level on the relative position of the shells \cite{kwon:prb1998}, we expect this effect to decrease strongly with the nanotube diameter. Our method is therefore suitable for the realistic DWNTs with diameters above 2nm \cite{hirahara:prb2006}.\\
When this method is applied to the DWNTs in the parallel magnetic field, we observe complex geometrical patterns developing in the DOS of the nanotubes. The most prominent ones are at energies inaccessible experimentally, but we find the effects of the intershell coupling also at the Fermi level. In a double-armchair DWNT we find the metallic character of the tube persisting also at $\phi> 0$, while without this coupling the system would become semiconducting immediately after switching on the field. In an incommensurate zigzag@armchair DWNT we find the band crossing of the zigzag shell shifted towards higher energies, while the band structure at the Fermi level in the armchair shell is almost unaffected by the intershell coupling.\\ 
The real-space methods of finding the spectrum of long commensurate DWNTs, where it is possible to define a common unit cell, are usually fast enough. In the case of incommensurate DWNTs the real space approach must be either to diagonalize the Hamiltonian of the whole DWNT, or to squeeze or stretch one of the shells so that they become commensurate and an approximate supercell can be found. The former is very costly in terms of computation time and memory, the second involves a deformation of the lattice of one or both shells. Solving the Schr\"{o}dinger equation in momentum space, as described here, allows us to use the selection rules and significantly reduce the size of the matrices to diagonalize. This method has been proven correct for short incommensurate nanotubes and for long commensurate DWNTs in a parallel magnetic field, where it gives the same results as those obtained in \cite{nemec:prb2006}. It may be a useful tool in investigating other properties of DWNTs.\\ 

\ack{
  The authors would like to thank G. Cuniberti
and N. Nemec for useful discussions. They acknowledge
  the support of DFG under the programs GRK 638 and SFB 689.
}

\section*{References}


\begin{thebibliography}{10}

\bibitem{saito:1998}
R.~Saito, G.~Dresselhaus, and M.~S. Dresselhaus.
\newblock {\em Physical Properties of Carbon Nanotubes}.
\newblock Imperial College Press, London, 1998.

\bibitem{book:2006}
A.~Loiseau, P.~Launois, P.~Petit, S.~Roche, and J.-P. Salvetat, editors.
\newblock {\em Understanding Carbon Nanotubes: From Basics to Applications}.
\newblock Springer, Berlin Heidelberg, 2006.

\bibitem{white:nature1998}
C.~T. White and T.~N. Todorov.
\newblock Carbon nanotubes as long ballistic conductors.
\newblock {\em Nature}, 393:240, 1998.

\bibitem{frank:science1998}
S.~Frank, P.~Poncharal, Z.~L. Wang, and W.A. de~Heer.
\newblock Carbon nanotube quantum resistors.
\newblock {\em Science}, 280:1744, 1998.

\bibitem{urbina:prl2003}
A.~Urbina, I.~Echeverria, A.~Perez-Garrido, A.~Diaz-Sanchez, and J.~Abellan.
\newblock Quantum conductance steps in solutions of multiwalled carbon
  nanotubes.
\newblock {\em Phys. Rev. Lett.}, 90(10):106603, 2003.

\bibitem{langer:prl1996}
L.~Langer, V.~Bayot, E.~Grivei, J.-P. Issi, J.~P. Heremans, C.~H. Olk,
  L.~Stockman, C.~van Haesendonck, and Y.~Bruynseraede.
\newblock Quantum transport in a multiwalled carbon nanotube.
\newblock {\em Phys. Rev. Lett.}, 76:479, 1996.

\bibitem{bachtold:nature1999}
A.~Bachtold, C.~Strunk, J.-P. Salvetat, J.-M. Bonard, L.~Forro, T.~Nussbaumer,
  and C.~Sch\"onenberger.
\newblock Aharonov-bohm oscillations in carbon nanotubes.
\newblock {\em Nature}, 397:673, 1999.

\bibitem{fujiwara:prb1999}
A.~Fujiwara, K.~Tomiyama, and H.~Suematsu.
\newblock Quantum interference of electrons in multiwall carbon nanotubes.
\newblock {\em Phys. Rev. B}, 60:13492, 1999.

\bibitem{collins:prl2001}
P.G. Collins, M.~Hersam, M.~Arnold, R.~Martel, and Ph. Avouris.
\newblock Current saturation and electrical breakdown in multiwalled carbon
  nanotubes.
\newblock {\em Phys. Rev. Lett.}, 86(14):3128--3131, 2001.

\bibitem{bourlon:prl2004}
B.~Bourlon, C.~Miko, L.~Forr{\'o}, DC~Glattli, and A.~Bachtold.
\newblock Determination of the intershell conductance in multiwalled carbon
  nanotubes.
\newblock {\em Phys. Rev. Lett.}, 93(17):176806, 2004.

\bibitem{charlier:prb1991}
J.-C. Charlier, X.~Gonze, and J.-P. Michenaud.
\newblock First-principles study of the electronic properties of graphite.
\newblock {\em Phys. Rev. B}, 43:4579, 1991.

\bibitem{misu:jpsj1979}
A.~Misu, E.E. Mendez, and M.S. Dresselhaus.
\newblock Near infrared reflectivity of graphite under hydrostatic pressure: 1.
  experiment.
\newblock {\em J. Phys. Soc. Jpn}, 47:199, 1979.

\bibitem{ohta:prl2007}
T.~Ohta, A.~Bostwick, J.L. McChesney, T.~Seyller, K.~Horn, and E.~Rotenberg.
\newblock Interlayer interaction and electronic screening in multilayer
  graphene investigated with angle-resolved photoemission spectroscopy.
\newblock {\em Phys. Rev. Lett.}, 98:206802, 2007.

\bibitem{charlier:prl1993}
J.~C. Charlier and J.~P. Michenaud.
\newblock Energetics of multilayered carbon tubules.
\newblock {\em Phys. Rev. Lett.}, 70:1858, 1993.

\bibitem{zettl:science2000}
J.~Cumings and A.~Zettl.
\newblock Low-friction nanoscale linear bearing realized from multiwall carbon
  nanotubes.
\newblock {\em Science}, 289:602, 2000.

\bibitem{saito:jap1993}
R.~Saito, G.~Dresselhaus, and M.~S. Dresselhaus.
\newblock Electronic structure of double-layer graphene tubules.
\newblock {\em J. Appl. Phys.}, 73:494, 1993.

\bibitem{kwon:prb1998}
Y.-K. Kwon and D.~Tomanek.
\newblock Electronic and structural properties of multiwall carbon nanotubes.
\newblock {\em Phys. Rev. B}, 58:R16001, 1998.

\bibitem{lin:physicae2006}
T.S. Li M.F.~Lin Y.H.~Ho, G.W.~Ho.
\newblock Electronic excitations of double-walled armchair carbon nanotubes.
\newblock {\em Physica E}, 32:569, 2006.

\bibitem{pudlak:condmat2007}
R.~Pincak M.~Pudlak.
\newblock Electronic properties of double-layer carbon nanotubes.
\newblock {\em cond-mat/0712.4346v1}, 2007.

\bibitem{triozon:prb2004}
Francois Triozon, Stephan Roche, Angel Rubio, and Didier Mayou.
\newblock Electrical transport in carbon nanotubes: Role of disorder and
  helical symmetries.
\newblock {\em Phys. Rev. B}, 69:121410, 2004.

\bibitem{ahn:prl2003}
K.-H. Ahn, Yong-Hyun Kim, J.~Wiersig, and K.~J. Chang.
\newblock Spectral correlation in incommensurate multiwalled carbon nanotubes.
\newblock {\em Phys. Rev. Lett.}, 90:026601, 2003.

\bibitem{uryu:prb2004}
S.~Uryu.
\newblock Electronic states and quantum transport in double-wall carbon
  nanotubes.
\newblock {\em Phys. Rev. B}, 69(7):075402, 2004.

\bibitem{wang:prl2005}
S.~Wang and M.~Grifoni.
\newblock Helicity and electron-correlation effects on transport properties of
  double-walled carbon nanotubes.
\newblock {\em Phys. Rev. Lett.}, 95(26):266802, 2005.

\bibitem{yoon:prb2002}
Y.-G. Yoon, P.~Delaney, and S.G. Louie.
\newblock Quantum conductance of multiwall carbon nanotubes.
\newblock {\em Phys. Rev. B}, 66:073407, 2002.

\bibitem{lunde:prb2005}
A.M. Lunde, K.~Flensberg, and A.P. Jauho.
\newblock {Intershell resistance in multiwall carbon nanotubes: A Coulomb drag
  study}.
\newblock {\em Physical Review B}, 71(12):125408, 2005.

\bibitem{wang:2006}
S.~Wang, M.~Grifoni, and S.~Roche.
\newblock Anomalous diffusion and elastic mean free path in disorder-free
  multiwalled carbon nanotubes.
\newblock {\em Phys. Rev. B}, 74(12):121407, 2006.

\bibitem{lambin:prb2000}
Ph. Lambin, V.~Meunier, and A.~Rubio.
\newblock Electronic structure of pholychiral carbon nanotubes.
\newblock {\em Phys. Rev. B}, 62:5129, 2000.

\bibitem{peierls:zphys1933}
R.~Peierls.
\newblock Zur theorie des diamagnetismus von leitungselektronen.
\newblock {\em Z. Phys.}, 80:763, 1933.

\bibitem{hofstadter:prb1976}
D.~R. Hofstadter.
\newblock Energy levels and wave functions of bloch electrons in rational and
  irrational magnetic fields.
\newblock {\em Phys. Rev. B}, 14:2239, 1976.

\bibitem{buttiker:prb1985}
M.~B\"{u}ttiker.
\newblock Small normal-metal loop coupled to an electron reservoir.
\newblock {\em Phys. Rev. B}, 32:1846, 1985.

\bibitem{gefen:prb1988}
H.-F. Cheung, Y.~Gefen, E.~K. Riedel, and W.-H. Shih.
\newblock Persistent currents in small one-dimensional metal rings.
\newblock {\em Phys. Rev. B}, 37:6050, 1988.

\bibitem{ajiki:jpsj1993a}
H.~Ajiki and T.~Ando.
\newblock Electronic states of carbon nanotubes.
\newblock {\em J. Phys. Soc. Jpn}, 62:1255, 1993.

\bibitem{mceuen:nature2004}
E.D. Minot, Y.~Yaish, V.~Sazonova, and P.~McEuen.
\newblock Determination of electron orbital magnetic moments in carbon
  nanotubes.
\newblock {\em Nature}, 428:536, 2004.

\bibitem{coskun:science2004}
U.C. Coskun, T.-Z. Wei, S.~Vishveshwara, P.M. Goldbart, and A.~Bezryadin.
\newblock h/e magnetic flux modulation of the energy gap in nanotube quantum
  dots.
\newblock {\em Science}, 304:1132, 2004.

\bibitem{strunk:sst2006}
C.~Strunk, B.~Stojetz, and S.~Roche.
\newblock Quantum interference in multiwall carbon nanotubes.
\newblock {\em Semicond. Sci. Technol.}, 21:38, 2006.

\bibitem{lassagne:prl2007}
B.~Lassagne, J-P. Cleuziou, S.~Nanot, W.~Escoffier, R.~Avriller, S.~Roche,
  L.~Forr\'{o}, B.~Raquet, and J.-M. Broto.
\newblock Aharonov-bohm conductance modulation in ballistic carbon nanotubes.
\newblock {\em Phys. Rev. Lett.}, 98:176802, 2007.

\bibitem{latge:carbon2007}
A.~Latg\'{e} and D.~Grimm.
\newblock Band-gap modulations of double-walled carbon nanotubes under an axial
  magnetic field.
\newblock {\em Carbon}, 45:1905, 2007.

\bibitem{lin:jpcm2008}
C.H. Lee, Y.C. Hsue, R.B. Chen, T.S. Li, and M.F. Lin.
\newblock Electronic structures of finite double-walled carbon nanotubes in a
  magnetic field.
\newblock {\em J.Phys.: Condens. Matter}, 20:75213, 2008.

\bibitem{nemec:prb2006}
N.~Nemec and G.~Cuniberti.
\newblock Hofstadter butterflies of carbon nanotubes: pseudofractality of the
  magnetoelectronic spectrum.
\newblock {\em Phys. Rev. B}, 74:165411, 2006.

\bibitem{charlier:prl2001}
J.-C. Charlier and G.-M. Rignanese.
\newblock Electronic structure of carbon nanocones.
\newblock {\em Phys. Rev. Lett.}, 86:5970, 2001.

\bibitem{chen:apl2006}
I-C. Chen, L.-H. Chen, X.-R. Ye, C.~Daraio, S.~Jin, C.A. Orme, A.~Quist, and
  R.~Lal.
\newblock Extremely sharp carbon nanocone probes for atomic force microscopy
  imaging.
\newblock {\em Appl. Phys. Lett.}, 88:153102, 2006.

\bibitem{ijima:nature1991}
S.~Ijima.
\newblock Helical microtubules of graphitic carbon.
\newblock {\em Nature}, 354:56, 1991.

\bibitem{maarouf:prb2000}
A.~A. Maarouf, C.~L. Kane, and E.~J. Mele.
\newblock Electronic structure of carbon nanotube ropes.
\newblock {\em Phys. Rev. B}, 61:11156, 2000.

\bibitem{kleiner:prb2001}
S.~Eggert A.~Kleiner.
\newblock Curvature, hybridization and stm images of carbon nanotubes.
\newblock {\em Phys. Rev. B}, 64:113402, 2001.

\bibitem{ding:jpcm2003}
J.X. Cao D.L. Wang Y. Tang Q.B.~Yang J.W.~Ding, X.H.~Han.
\newblock Curvature and strain effects on electronic properties of single-wall
  carbon nanotubes.
\newblock {\em J. Phys.: Consens. Matter}, 15:439, 2003.

\bibitem{mayer:carbon2004}
A.~Mayer.
\newblock Band structure and transport properties of carbon nanotubes using a
  local pseudopotential and a transfer-matrix technique.
\newblock {\em Carbon}, 42:2057, 2004.

\bibitem{nemec:phd2007}
N.~Nemec.
\newblock {\em Quantum transport in carbon-based nanostructures}.
\newblock PhD thesis, University of Regensburg, 2007.

\bibitem{ajiki:jpsj1993b}
H.~Ajiki and T.~Ando.
\newblock Magnetic properties of carbon nanotubes.
\newblock {\em J. Phys. Soc. Jpn}, 62:2470, 1993.

\bibitem{hirahara:prb2006}
K.~Hirahara, M.~Kociak, S.~Bandow, T.~Nakahira, K.~Itoh, Y.~Saito, and
  S.~Iijima.
\newblock Chirality correlation in double-wall carbon nanotubes as studied by
  electron diffraction.
\newblock {\em Phys. Rev. B}, 73(19):195420, 2006.

\end{thebibliography}

\appendix

\section{Derivation of the intershell coupling matrix elements}
\label{sec:intershell_coupling}

In this appendix we give an explicit derivation of the elements of the
intershell coupling matrix, \eref{eq:effective-coupling}.
By expressing the Hamiltonian in the basis of plane waves in each
shell, the intershell coupling Hamiltonian $H_t$ between two shells $a$
and $b$ can be written as
\begin{eqnarray*}
  H_t &= \sum_{ij\sigma} t_{\ve{r}_{ai},\ve{r}_{bj}} \cre{c}{ai\sigma}\ann{c}{bj\sigma} +
  \mathrm{H.c.}  \\
    &= \frac{1}{\sqrt{N_aN_b}} \sum_{ij\sigma} 
      \sum_{\ve{k}_a\ve{k}_b p_a p_b} 
e^{-i\ve{k}\cdot\ve{r}_{ai} + i\ve{k}\cdot\ve{r}_{bj}}
t_{\ve{r}_{ai},\ve{r}_{bj}} 
      \cre{c}{ap_a\ve{k}_a\sigma}\ann{c}{bp_b\ve{k}_b\sigma} + \mathrm{H.c},
\end{eqnarray*}
where $p_a, p_b = \pm$ are the indices for the two interpenetrating
sublattices, $N_a$ and $N_b$ are the number of graphene unit cells on shell
$a$ and $b$, respectively. A carbon atom can be found in graphene at
the position $\ve{r} = \ve{R} + \ve{\rho} + p\ve{\tau}$, where $\ve{R}$ is
a graphene lattice vector and $\ve{\rho}$ and $\ve{\tau}$ are two vectors that
specify the atom position, cf. \fref{fig:dwnt}(a). The sum over
all the lattice sites can be carried out as
\begin{eqnarray*}
\fl    \frac{1}{\sqrt{N_aN_b}}& \sum_{ij}
e^{-i\ve{k}\cdot\ve{r}_{ai} + i\ve{k}\cdot\ve{r}_{bj}}
t_{\ve{r}_{ai},\ve{r}_{bj}}  \\
\fl & = 
    \frac{1}{\sqrt{N_aN_b}} \sum_{\ve{R}_a\ve{R}_bp_ap_b} 
e^{-i\ve{k}\cdot\ve{r}_{ai} + i\ve{k}\cdot\ve{r}_{bj}}
t_{\ve{r}_{ai},\ve{r}_{bj}} \Big\vert_{
{\ve{r}_a = \ve{R}_a + \brho + p_a\btau}, 
{\ve{r}_b = \ve{R}_b + \brho + p_b\btau}}  \\
\fl & = \frac{1}{\sqrt{N_aN_b}} \sum_{p_ap_b} 
 \int d\ve{r}_a d\ve{r}_b 
 \sum_{\ve{R}_a\ve{R}_b}\delta(\ve{r}_a - (\ve{R}_a + \brho + p_a\btau)) \\
\fl & \times \delta(\ve{r}_b - (\ve{R}_b + \brho + p_b\btau))
e^{-i\ve{k}\cdot\ve{r}_{ai} + i\ve{k}\cdot\ve{r}_{bj}}
t_{\ve{r}_{ai},\ve{r}_{bj}} \\
\fl & = \frac{1}{\sqrt{N_aN_b}A_{\mathrm{cell}}^2}
 \sum_{\ve{G}_a\ve{G}_bp_ap_b}
e^{i \ve{G}_a\cdot(\brho + p_a\btau) - i \ve{G}_b \cdot
   (\brho + p_b\btau)} \\
\fl & \times \int d\ve{r}_a d\ve{r}_b 
e^{-i(\ve{k}_a+\ve{G}_a)\cdot\ve{r}_a + i (\ve{k}_b
   + \ve{G}_b)\cdot\ve{r}_b } 
 t_{\ve{r}_{ai},\ve{r}_{bj}} \;,
\end{eqnarray*}
where we have replaced the sum over the lattice vectors $\ve{R}$ by
the sum over the reciprocal lattice vector $\ve{G}$ as
\begin{equation*}
  \sum_{\ve{R}} 
 \delta(\ve{r} - (\ve{R} + \brho + p\btau)) 
=  \frac{1}{A_{\mathrm{cell}}} \sum_{\ve{G}} e^{-i\ve{G}\cdot(\ve{r} - \brho
    - p \btau)}
\end{equation*}
with $A_{\mathrm{cell}}$ the area of a graphene unit cell. 
Thus we obtain the expression of the elements of the intershell
coupling matrix, \eref{eq:effective-coupling}.

\end{document}